%% file: tech-report.tex
\documentclass[11pt]{article}
\usepackage[margin=1in]{geometry}
\usepackage{graphicx}
\usepackage{paralist}
\usepackage{algorithm,algorithmic}

\newif\ifCONF
\CONFfalse

\usepackage{hyperref}
\usepackage{amsmath,amsfonts}
\usepackage{multirow, multicol}
\usepackage{pgfplots,diagbox,tikz,url}
\usepackage{subcaption}
\newtheorem{theorem}{Theorem}
\newtheorem{claim}{Claim}
\newtheorem{definition}{Definition}

\long\def\ignore#1{}
\usepackage{subfiles} 

\title{Secure and Accurate Summation of Many Floating-Point Numbers}
\author{Marina Blanton\\
  Department of Computer Science and Engineering\\
  University at Buffalo\\
 {mblanton@buffalo.edu}
\and 
Michael T. Goodrich\\
Department of Computer Science\\
University of California, Irvine\\
{goodrich@acm.org}
\and
Chen Yuan\thanks{Majority of the work was performed while at the University at Buffalo.}\\
Meta Platform Inc.\\
{chenyc@meta.com}
}
\date{}

\begin{document}
\maketitle

\begin{abstract}
Motivated by the importance of floating-point computations, we study the problem of securely and accurately summing many floating-point numbers. Prior work has focused on security absent accuracy or accuracy absent security, whereas our approach achieves both of them. Specifically, we show how to implement floating-point superaccumulators using secure multi-party computation techniques, so that a number of participants holding secret shares of floating-point numbers can accurately compute their sum while keeping the individual values private.
\end{abstract}

\section{Introduction}
\subfile{sections/introduction}

\section{Floating-Point Summation Construction}

\subfile{sections/construction}

\subfile{sections/buildingblock}

\subfile{sections/large-precision-construction}

\newcommand{\perftables}{
\begin{table*}[t]
\ifCONF \else {\small \setlength{\tabcolsep}{0.9ex}\fi
    \begin{subtable}[h]{\textwidth}
    \centering
  \begin{tabular}{|l|r|r|r|r|r|r|r|r||r|r|r|r|r|r|r|r|r|r|} 
  \hline
    \multirow{3}{*}{\hspace{0.05in}Prot.} & \multicolumn{16}{c|}{Input size}\\ \cline{2-17}
    & \multicolumn{8}{c||}{$w=16$} & \multicolumn{8}{c|}{$w=32$} \\ \cline{2-17}
    & $2^4$ & $2^6$ & $2^8$ & $2^{10}$ & $2^{12}$ & $2^{14}$ & $2^{16}$ & $2^{18}$ & $2^4$ & $2^6$ & $2^8$ & $2^{10}$ & $2^{12}$ & $2^{14}$ & $2^{16}$ & $2^{18}$ \\ \hline
    ${\sf FL2SA}$  & 6.81 & 9.06 & 16.9   & 45.4 & 136 & 529 & 2160 & 8403 & 6.02 & 8.13 & 17.5 & 46.4 & 142 & 585 & 2324 & 9036 \\ \hline
    ${\sf SASum}$ & 3.22 & 3.14 & 3.81    & 3.68  & 3.56  & 4.37  & 20.5 & 85.1 & 3.1 & 3.17  & 3.68  & 3.71  & 3.89  & 4.17  & 18.4 & 79.2 \\ \hline
    ${\sf SA2FL}$ & 6.82 & 6.75 & 6.74     & 6.67  & 6.48  & 6.74  & 6.87 & 6.71 & 7.83 & 7.84  & 7.84  & 7.86  & 7.74  & 7.89  & 7.91 & 7.74\\ \hline
    Total & 16.8 & 18.9 & 27.4   & 55.7 & 146 & 540 & 2187 & 8495 & 16.9 & 19.1  & 28.7 & 57.9  & 154   & 598   & 2351 & 9124 \\ \hline
  \end{tabular}
  \caption{Single floating-point precision.}
  \end{subtable}

  \begin{subtable}[h]{\textwidth} 
  \ifCONF \else \setlength{\tabcolsep}{0.8ex} \fi
  \centering
  \begin{tabular}{|l|r|r|r|r|r|r|r|r||r|r|r|r|r|r|r|r|r|r|} \hline
    \multirow{3}{*}{\hspace{0.05in}Prot.} & \multicolumn{16}{c|}{Input size}\\ \cline{2-17}
    & \multicolumn{8}{c||}{$w=16$} & \multicolumn{8}{c|}{$w=32$} \\ \cline{2-17}
    & $2^4$ & $2^6$ & $2^8$ & $2^{10}$ & $2^{12}$ & $2^{14}$ & $2^{16}$ & $2^{18}$ & $2^4$ & $2^6$ & $2^8$ & $2^{10}$ & $2^{12}$ & $2^{14}$ & $2^{16}$ & $2^{18}$ \\ \hline
    ${\sf FL2SA}$  & 9.09     & 14.3  & 36.2 & 114   & 413   & 1668 & 6688 & 24805  & 9.32 & 14.3 & 32.9    & 106.9 & 384& 1517 & 6247 & 23486 \\ \hline
    ${\sf SASum}$ & 4.43      & 4.87  & 4.91  & 4.93  & 6.29 & 10.4  & 17.4 & 57.4 & 4.78 & 4.87   & 5.17  & 5.07 & 6.39  & 9.71 & 14.7 & 45.0\\ \hline
    ${\sf SA2FL}$ & 8.78      & 8.49  & 8.41  & 8.31  & 8.12  & 8.22  & 8.25 & 8.24 & 9.31  & 9.14  & 9.13 & 8.97   & 9.04  & 8.87  & 8.71 & 9.23\\ \hline
    Total  & 22.3      & 27.7  & 49.5  & 127   & 427   & 1687  & 6714 & 24871   & 23.4  & 28.3 & 47.2 & 121    & 399   & 1536  & 6271 & 23540   \\ \hline
  \end{tabular}
  \caption{Double floating-point precision.}
  \end{subtable}
\ifCONF \else } \fi
\caption{Performance $\sf FLSum$ in ms.} \label{tab:perf}    
\end{table*}}

\subfile{sections/performance}

\section{Conclusions}

The goal of this work is to develop secure protocols for accurate summation of many floating-point values that avoid round-off errors of conventional floating-point addition. Our solution uses the notion of a superaccumulator and the computation proceeds by converting floating-point inputs into superaccumulator representation, performing exact summation, and converting the computed result back to a floating-point value. Despite providing higher accuracy, we demonstrate that our solution outperforms state-of-the-art secure floating-point summation.

\section*{Acknowledgements}

This work was supported in part by NSF grants 2213057 and 2212129. Any opinions, findings, and conclusions or recommendations expressed in this publication are those of the authors and do not necessarily reflect the views of the funding sources. The authors would also like to thank anonymous reviewers, Alessandro Baccarini for help with running the experiments, and Haodi Wang for detecting multiple typos.

\bibliographystyle{plain}
\bibliography{floating-reference}

\end{document}

%% file: sections/introduction.tex
Floating-point numbers are the most widely used 
data type for approximating real numbers with a 
wide variety of applications; see, e.g., \cite{Goldberg:1991:CSK:103162.103163,muller2009handbook,Wang-float}.
A (radix-2) floating-point number $x$
is a tuple of integers $(b,v,p)$ such that
\begin{equation}
x = (-1)^b \times (1+ 2^{-m}v) \times 2^{p-2^{e-1}-1},
\label{eq:float}
\end{equation}
where $b\in\{0,1\}$ is a \emph{sign bit},
$v$ is the $m$-bit \emph{mantissa} (which is also known as the \emph{significand}), and $p$ is the $e$-bit \emph{exponent}. 

A well-known issue with floating-point arithmetic is that it is not exact.
For example,
it is known that
summing two floating point numbers can have a roundoff error and these
roundoff errors can propagate and even become larger than a computed result
when performing a sequence of many floating-point additions.
For example, floating-point addition is not 
associative~\cite{Knuth:1997:ACP:270146}.

Floating-point arithmetic has applications in many areas including  
medicine, defense, economics, and physics simulation (e.g., 
in the NVIDIA Omniverse~\cite{hummel2019leveraging}).
Thus, there is considerable 
need in computing sums of many floating-point numbers as accurately as possible.
For example, the accuracy of any computation that involves 
high-dimensional dot products or matrix multiplications, 
such as in machine-learning 
(see, e.g., \cite{dot-product,fasi2021numerical}),
depends on the accuracy of computing the sum of many floating-point numbers.
Similarly, computations in computational geometry involve computing
determinants, whose accuracy also depends on computing the sum of
many floating-point numbers; see, e.g., \cite{demmel,she1997,shewchuk}.

In addition, the fact that floating-point addition is not associative 
presents problems related to the reproducibility of computations;
see, e.g.,~\cite{doi:10.1137/S1064827502407627,demmel,CDG+14,DN6875899}.
For example, a secure contract involving the summation of floating-point
numbers may need to be verified after it has been signed.
And if this depends on the summation of floating-point values, performing the summation
on different computers could result in different outcomes, which could cause
participants to reject an otherwise valid digital contract.

Competing with this issue is that some applications of floating-point
arithmetic have computer-security requirements, including integrity,
confidentiality, and privacy.
For example, computing the probability of satellites colliding could
involve security and privacy considerations when the satellites
belong to competing companies or adversarial nation-states, e.g.,
see~\cite{kam2015}.
Thus, there is a need for protocols for computing sums of 
many floating-point numbers as securely as possible.
This holds for other domains where computation on private data is performed using floating-point arithmetic including applications in medicine and privacy-preserving training of machine learning models on distributed sensitive data. 

In spite of the importance of accuracy
and security for summing floating-point numbers,
we are not aware of any prior work that simultaneously achieves both
accuracy and security for summing many floating-point numbers.
As we review below, there is considerable prior work on 
methods for accurately summing many floating-point numbers, but the methods
used do not lend themselves to transformations into secure computations.
Likewise, as we also review below, there is considerable prior work on
securely computing sums of pairs of floating-point numbers, but these prior
methods do not consider the propagation of roundoff errors and can lead to inaccurate results for summing many floating-point numbers. Such inaccuracies can arise after adding numbers of significantly different magnitudes, where the values of the largest magnitude have opposite signs and significantly exceed other summation operands. Adding the values one at a time using floating-point addition can therefore leave us with noise, while implementing addition exactly will retain the necessary number of summation bits.
Thus, in this paper, we are interested in methods for summing many
floating-point numbers that are both secure and accurate.

\medskip \noindent \textbf{Related Prior Work.}
Neal~\cite{Neal15a} describes algorithms using a number
representation called a \emph{superaccumulator} to exactly
sum $n$ floating point numbers, which is then converted
to a faithfully-rounded floating-point number.
Unfortunately, while Neal's superaccumulator representation
reduces carry-bit propagation, it does not eliminate it, as is needed 
for the purposes of this work.
A similar idea has been used in ExBLAS~\cite{CDG+14}, an open source
library for floating point computations.
Shewchuck~\cite{shewchuk} describes an alternative
representation for exactly representing intermediate results of
floating-point arithmetic, but the method also does not eliminate
carry-bit propagation in summations; hence, it also does not satisfy our accuracy constraints.
In addition to these solutions, there are a number of 
adaptive methods for exactly summing $n$ floating point numbers
using various other data structures for representing intermediate results,
which do not consider the security or privacy of the data.
Further, these methods, which include ExBLAS~\cite{CDG+14} and algorithms by Zhu and Hayes~\cite{doi:10.1137/070710020,Zhu:2010:A9O}, Demmel and Hida~\cite{doi:10.1137/S1064827502407627,demmel}, Rump {\it et al.}~\cite{doi:10.1137/050645671}, Priest~\cite{p145549}, Malcolm~\cite{Malcolm:1971},
Leuprecht and Oberaigner~\cite{par82},
Kadric {\it et al.}~\cite{KGD6545903},
and Demmel and Nguyen~\cite{DN6875899},
are not amenable to conversion to secure protocols with few rounds.


While integer arithmetic in secure multi-party computation has 
been extensively investigated, secure floating-point 
arithmetic has only gradually attracted attention in the last decade.
Catrina and Saxena~\cite{cat2010f} extended secure computation from 
integer pairwise arithmetic to fixed-point pairwise arithmetic 
and applied it to linear programming \cite{cat2010s}.
Franz and Katzenbeisser~\cite{fra2011} proposed a 
solution, based on homomorphic encryption and garbled circuits, 
for floating-point pairwise operations in the 
two-party setting with no implementation or performance results.
Aliasgari {\it et al.}~\cite{ali2013} designed a set of protocols for 
basic floating-point operations based on Shamir secret sharing 
and developed several advanced operations such as logarithm, 
square root and exponentiation of floating-point numbers. 
Their solution was improved and extended for other settings and applications \cite{ali2017,bay2017,sas2020,kam2015} later.
Dimitrov et al.~\cite{dim2016} proposed two sets of protocols using new representations to improve efficiency, but did not follow the IEEE 754 standard representation. Archer et al.~\cite{arc21} measure performance of floating-point operations in different instantiations using a varying number of computation participants and corruption thresholds. Rathee et al.~\cite{SecFloat} design secure protocols in the two-party setting and exactly follow the IEEE standard rounding procedure. 
In addition to the above works on improving efficiency of unary/binary floating-point operations, 
Catrina~\cite{cat2020,cat2020p,cat2021} proposed and improved several 
multi-operand operations such as sum, dot-product, and polynomial evaluation.
Nevertheless, because their solutions are still based on 
traditional floating-point pairwise addition,
round-off errors accumulate inevitably in each addition operation.

\medskip \noindent \textbf{Our Results.}
In this paper, we
develop new secure protocols for summing many floating-point numbers that outperforms other approaches. 
We design a superaccumulator-based solution that privately and 
accurately calculates summations of many private arbitrary-precision 
floating-point numbers, and we
empirically evaluate the performance of our solution on varying input sizes 
and precision. Unlike standard floating-point addition, our approach performs summation exactly without introducing round-off errors. 

Our supperaccumulator-based approach and most of the protocols we develop can be instantiated with building blocks based on secret sharing in different settings, including computation with or without honest majority and semi-honest and malicious adversarial models. Some of the design choices are made in favor of reducing communication and one efficient low-level building block, conversion shares of a bit from binary to arithmetic sharing, is in the three-party setting with honest majority based on replicated secret sharing in the semi-honest model (as defined below). We implement the construction in that setting and show that its runtime is faster than the state of the art implementing floating-point operations~\cite{cat2020p,SecFloat}. Thus, we are able to implement exact addition while simultaneously improving performance.

%% file: sections/construction.tex
\subsection{The Expand-and-Sum Solution}
\label{sec:brute}
There is a simple na{\"\i}ve solution for exactly summing a set of $n$ 
floating-point numbers, $\{x_1,x_2,\ldots,x_n\}$,
which we refer to as the \emph{expand-and-sum} solution.
It is reasonable for low-precision floating-point representations and is given as Algorithm~\ref{alg:expand}.
\begin{algorithm}[t] 
  \caption{$s \leftarrow {\sf ExpandAndSum}(x_1,x_2,\ldots,x_n)$}
  \label{alg:expand}
  \begin{algorithmic}[1]
     \FOR{$i = 1, \ldots, n$}
       \STATE $y_i \leftarrow {\sf ConvertToInt}(x_i)$;
     \ENDFOR
     \STATE $v \leftarrow \sum_{i=1}^n y_i$; ~// exact addition
     \STATE $s \leftarrow {\sf ConvertToFloat}(v)$;
     \RETURN $s$;
  \end{algorithmic}
\end{algorithm}
That is, for each floating-point number $x_i$, we convert 
the representation of $x_i$ into an integer $y_i$, 
with as many bits as is possible based on the floating-point type 
being used for the $x_i$s.
Then we sum these values exactly using integer addition 
and convert the result back into a floating-point number.

The $y_i$s would have the following sizes 
based on the IEEE 754 formats:
\begin{itemize}
\item \emph{Half}: a half-precision floating-point number
in the IEEE 754 format has 1 sign bit, a 5-bit exponent, and a 10-bit mantissa. Thus, representing this as an integer requires $1+2^5+10= 43$ bits.

\item \emph{Single}: a single-precision floating-point number has 1 sign bit, an 8-bit exponent, and a 23-bit mantissa. Thus, representing this as an integer requires $1+2^8+23 = 280$ bits.

\item \emph{Double}: a double-precision floating-point number has 1 sign bit, an 11-bit exponent, and a 52-bit mantissa. Thus, representing this as an integer requires $1+2^{11}+52 = 2,101$ bits.

\item \emph{Quad}: a quad-precision floating-point number has 1 sign bit, a 15-bit exponent, and a 112-bit mantissa. Thus, representing this as an integer requires $1+2^{15}+112 = 32,881$ bits.
\end{itemize}
Further, there are also even higher-precision 
floating-point representations, which would require even more bits to represent as fixed-precision or integer numbers; see,  e.g.,~\cite{apfloat,GMP,leda,mpfr,mpfr07}.
Implementing a summation using this representation would involve performing many operations on very large numbers using secure multi-party computation techniques, thus degrading performance.
Of course, applications with high-precision floating-point numbers are likely to be applications that require accurate summations; hence, we desire solutions that can work efficiently for such applications without requiring ways of summing very large integers.
In particular, summing very large integers requires  techniques for dealing with cascading carry bits during the summations, and performing all these operations securely is challenging for very large integers.
Thus, 
we consider this expand-and-sum approach for summing $n$ floating-point numbers as integers to be limited to low-precision floating-point representations.

\subsection{Superaccumulators}
\label{sec:superaccu}

An alternative approach, which is better suited for use with conventional secure addition when applied to high-precision
floating-point formats, is to use a \emph{superaccumulator} to represent floating-point summands, e.g., see~\cite{collange2014full,CDG+14,Neal15a}.
This approach also uses integer arithmetic, but with much smaller integers. More importantly, 
it can be implemented to avoid cascading carry-bit propagation. 

In a superaccumulator, instead of representing a floating-point number as a single expanded (very-large) integer, we represent that integer as a sum of small components maintained separately.
That is, we represent the expanded integer $y$, 
corresponding to a floating-point number $x$, as a vector of $2w$-bit integers $\langle y_{\alpha},y_{\alpha-1},\ldots,y_1 \rangle$, where $y = \sum_{i=1}^{\alpha} (2^w)^{i-1} y_i$
and $\alpha = \lceil \frac{2^e+m}{w}\rceil$, so that we cover all possible
exponent values.
Also, note that if we convert a floating-point number to a superaccumulator,
then at most $\beta = \lceil \frac{m+1}{w} \rceil + 1 $ 
of the entries will be non-zero.
We can choose $w$ based on the underlying mechanism for achieving security and privacy. For example, if we want to use built-in 64-bit integer addition, we can choose $w$ to be~32.

In addition, we say that $s$ is \emph{regularized} 
if $-2^w < y_i < 2^w$ for $i=1, \ldots, \alpha$.
At a high level,
in our scheme, we start with a regularized representation for 
each floating-point number $x_i$, and then we perform summations
on an element-by-element basis. Finally, we regularize the partial sums
by shifting ``carry'' values to neighboring elements.
As we show, this approach allows us to prevent these carry values
from propagating in a cascading fashion after performing a group of sums, 
which allows us to achieve efficiency
for our secure summation protocols.

\ignore{
Our method for converting a floating-point value $x_i$ to a superaccumulator $y_i$ is given in Algorithm~\ref{alg:super}, where $x_i$ is represented using a sign bit $b_i$, an $m$-bit mantissa $v_i$, and an $e$-bit exponent $p_i$.
\begin{algorithm}[t]
  \caption{$y_i \leftarrow {\sf SuperAccumulator}(x_i = \langle b_i, v_i, p_i \rangle)$}
  \label{alg:super}
  \begin{algorithmic}[1]
   \FOR{\textbf{each} $i = l, \ldots, h$}
        \STATE $y_{i,j} = 0$;
    \ENDFOR
    \STATE xxx DO WE WANT TO ASSUME THAT WE JUST SPLIT INTO 2 COMPONENTS?
     \RETURN $s$;
  \end{algorithmic}
\end{algorithm}
To convert a floating-point number $x_i$, into a regularized superaccumulator, we need to shift the bits of $x_i$ appropriately, split the binary representation into at most 2 $y_j$s, and pad the rest of the $y_j$ values to be $0$.

xxx This assumes that we just have 2 $y_j$'s that are non-zero!!! xxx

xxx converting back to floating-point xxx

To convert a regularized superaccumulator $s = \sum_{j=l}^{h} 2^{w_j} y_j$ to a floating-point number $x$, we find the largest $j$ such that $y_j\not=0$ and sum the next $2^{w_j} y_j$ values.

xxx performing a summation xxx
}

Suppose we are given $n$ floating-point numbers, $\{x_1,x_2,\ldots,x_n\}$,
each represented as a regularized superaccumulator
$x_i = \sum_{j=1}^{\alpha} (2^w)^{j-1} y_{i,j}$.
Further, suppose $n\le 2^{w-2}$.
We sum all the $x_i$'s by 
\begin{itemize}
\item first summing the corresponding terms,
$s_j = \sum_{i=1}^{n} y_{i,j}$,
\item then splitting the binary representation of each $s_j$ into $c_{j+1}$ and
$r_j$, so that 
$s_j = c_{j+1}2^{w-1} + r_j$,
where $-2^{w-1}<r_j<2^{w-1}$,
\item and lastly, updating each $s_j$ as
$s_j \leftarrow r_j + c_j$,
for $j=1,\ldots,n$.
\end{itemize}
As we show, because of the way that we regularize superaccumulators,
the ``carry'' values, $c_j$, 
will not propagate in a cascading way, and the result of the 
above summation will be regularized. This
allows us to complete the sum in a single communication round.

Further, for practical values of $w$, the constraint
that $n \le 2^{w-2}$ is not restrictive.
For example, if $w=32$, this implies we can sum up to one billion
floating-point numbers
in a single communication round.
Thus, to sum larger groups of numbers, we can
group the summations in a tree where each internal node
has $2^{w-2}$ children, and perform the sums in a bottom-up fashion.
The important property, though, is that performing the above approach
of summing
$n \le 2^{w-2}$ regularized superaccumulators and then adding the
carry values, $c_j$ (some of which may be negative), 
to the neighboring element will result in 
a regularized superaccumulator.
The following theorem establishes this property.

\begin{theorem}
If $n\le 2^{w-2}$, then summing $n$
regularized superaccumulators
using the above algorithm will produce a regularized result.
\end{theorem}

\ifCONF \begin{proof}
\else \noindent \textbf{Proof}
\fi
Let $x_1,x_2,\ldots,x_n$ be the set of input superaccumulators to sum,
where $n\le 2^{w-2}$ and
$x_i = \sum_{j=1}^{\alpha} (2^w)^{j-1} y_{i,j}$
for $i=1,2,\ldots,n$.
Recall that we sum all the $x_i$s by summing the corresponding terms, i.e., 
$s_j = \sum_{i=1}^{n} y_{i,j}$.
Since each $x_i$ is regularized, $-2^w < y_{i,j} < 2^w$ for all $i,j$.
Thus, $-2^w n < s_j < 2^w n$ for all $j$;
and hence, $-2^{2w-2} < s_j < 2^{2w-2}$ since $n\le 2^{w-2}$.

Recall that we  split the binary representation of each $s_j$ into $c_{j+1}$ and $r_j$, so that 
$s_j = c_{j+1}2^{w-1} + r_j$,
where $-2^{w-1}<r_j<2^{w-1}$.
Thus,
\begin{center}
$s_j = c_{j+1}2^{w-1} + r_j < c_{j+1}2^{w-1} + 2^{w-1} =  
(c_{j+1}+1)2^{w-1} < 2^{2w-2}$ and\\[0.05in]
$s_j = c_{j+1}2^{w-1} + r_j > c_{j+1}2^{w-1} - 2^{w-1} =  
(c_{j+1}-1)2^{w-1} > -2^{2w-2}.$
\end{center}
Therefore,
$-2^{w-1}+1 < c_{j+1} < 2^{w-1} - 1$
for each $j$.
So, when we update each $s_j$ as 
$s_j \leftarrow r_j + c_j$,
then
\begin{center}
$s_j = r_j + c_j < 2^{w-1} + 2^{w-1} -1 = 2^w -1$ and\\[0.05in]
$s_j = r_j + c_j > -2^{w-1} - 2^{w-1} +1 = -2^w +1.$
\end{center}
Therefore, the result is regularized.
\ifCONF \end{proof}
\else \hfill $\Box$
\fi

%% file: sections/buildingblock.tex
\section{Secure Computation Preliminaries}

\subsection{Security Setting}
We use a conventional secure multi-party setting with $N$ parties running the computation, $t$ of which can be corrupt. Given a function $f$ to be evaluated, the computational parties securely evaluate it on private data such that no information about the private inputs, or information derived from the private inputs, is revealed. More formally, a standard security definition requires that the view of the participants during the computation is indistinguishable from a simulated view generated without access to any private data. 

Most of the protocols developed in this work can be instantiated in different adversarial models, but our implementation and one low-level building block are 
in the semi-honest model, in which the participating parties are expected to follow the computation, but might try to learn additional information from what they observe during the computation.
Then the security requirement is that any coalition of at most $t$ conspiring computational parties is unable to learn any information about private data that the computation handles. Achieving security in the semi-honest setting first is also important if one wants to have stronger security guarantees, and many of the protocols developed in this work would also be secure in the malicious model when instantiated with stronger building blocks.

\begin{definition} \label{def:sec}
Let parties $P_1, {\ldots}, P_N$ engage in a protocol $\Pi$ that computes function $f({\sf in}_1, {\ldots}, {\sf in}_N) = ({\sf out}_1, {\ldots}, $ $ {\sf out}_N)$, where ${\sf in}_i$ and ${\sf out}_i$ denote the input and output of party $P_i$, respectively. Let $\mathrm{VIEW}_\Pi(P_i)$ denote the view of participant $P_i$ during the execution of protocol $\Pi$. More precisely, $P_i$'s view is formed by its input and internal random coin tosses $r_i$, as well as messages $m_1, {\ldots}, m_k$ passed between the parties during protocol execution: $\mathrm{VIEW}_{\Pi}(P_i) = ({\sf in}_i, r_i, m_1, {\ldots}, m_k).$ Let $I = \{P_{i_1}, P_{i_2}, {\ldots}, P_{i_t}\}$ denote a subset of the participants for $t < N$, $\mathrm{VIEW}_\Pi(I)$ denote the combined view of participants in $I$ during the execution of protocol $\Pi$ (i.e., the union of the views of the participants in $I$), and $f_I({\sf in}_1, \ldots, {\sf in}_N)$ denote the projection of $f({\sf in}_1, \ldots, {\sf in}_N)$ on the coordinates in $I$ (i.e., $f_I({\sf in}_1, \ldots, {\sf in}_N)$ consists of the $i_1$th, $\ldots$, $i_t$th element that $f({\sf in}_1, \ldots, {\sf in}_N)$ outputs).  We say that protocol $\Pi$ is $t$-private in the presence of semi-honest adversaries if for each coalition of size at most $t$ there exists a probabilistic polynomial time (PPT) simulator $S_I$ such that $\{S_I({\sf in}_I, f_I({\sf in}_1, \ldots, {\sf in}_n)), f({\sf in}_1, {\ldots}, {\sf in}_n)\} \equiv \{\mathrm{VIEW}_\Pi(I), ({\sf out}_1, \ldots, {\sf out}_n)\},$ where ${\sf in}_I = \bigcup_{P_i \in I} \{{\sf in}_i\}$ and $\equiv$ denotes computational or statistical indistinguishability.
\end{definition}

The focus of this work is on precise (privacy-preserving) floating-point summation, and this operation is typically a part of a larger computation. For that reason, the inputs into the summation would be the result of other computations on private data. Therefore, we assume that the inputs into the summation are not known by the computational parties and are instead entered into the computation in a privacy-preserving form. Similarly, the output of the summation can be used for further computation and is not disclosed to the parties. In other words, we are developing a building block that can be used in other computations, where the computational parties are given privacy-preserving representation of the inputs, jointly produce a privacy-preserving representation of the output, and must not learn any information about the values they handle. This permits our solution to be used in any higher-level computation and abstracts the setting from the way the inputs are entered into the computation (which can come from the computational parties themselves or external input providers).  

In our solution, we heavily rely on the fact that composition of secure building blocks is also secure. As part of this work, we develop several new building blocks to enable the functionality we want to support.

\subsection{Secret Sharing}
To realize secure computation, we utilize $(N, t)$-threshold linear secret sharing. Secret sharing offers efficiency due to the information-theoretic nature of the techniques and consequently the ability to operate over a small field or ring. Many of the protocols developed in this work can be realized using any suitable type of secret sharing (e.g,. with or without honest majority and in the semi-honest or malicious settings) and by $[x]$ we denote a secret-shared representation of value $x$, which is an element of the underlying field or ring. The expected properties are that (i) each of the $N$ computational parties $P_i$ holds its own share 
such that any combination of $t$ shares reveals no information about $x$ and (ii) a linear combination of secret-shared values can be computed by each party locally on its shares. SPD$\mathbb{Z}_{2^k}$~\cite{spdz2k} is one example of a suitable framework.

For performance reasons, many recent publications utilize computation over ring $\mathbb{Z}_{2^k}$ for some $k \ge 1$, which permits the use of native CPU instructions for performing ring operations. This is also the setting that we utilize for our experiments and use to inform certain protocol optimizations. Conventional techniques such as Shamir secret sharing~\cite{sha1979} cannot operate over $\mathbb{Z}_{2^k}$ and thus we rely on replicated secret sharing~\cite{ito1987} with a small number of parties. Specifically, we use the setting with honest majority, i.e., where $t < N/2$, and are primarily interested in the three-party setting, i.e., $N=3$. 
All parties $P_1, \ldots, P_N$ are assumed to be connected by pair-wise secure authenticated channels.

There is a need to secret share both positive and negative integers and the space is used to naturally represent all values as non-negative ring/field elements. In that case, the most significant bit of the representation determines the sign.  

For efficiency reasons, portions of the computation proceed on secret shared values set up over a different ring, most commonly $\mathbb{Z}_2$. Thus, we use notation $[x]_\ell$ to denote secret sharing over $\mathbb{Z}_{2^\ell}$ when $\ell$ differs from the default $k$.

\subsection{Building Blocks}
In a linear secret sharing scheme, a linear combination of secret-shared values can be performed locally on the shares without communication. This includes addition, subtraction, and multiplication by a known element. Multiplication of secret-shared values requires communication and the cost varies based on the setting. We use the multiplication protocol from \cite{bac2020} that works with any number of parties in the honest majority setting and communicates only one element in one round in the three-party setting, i.e., when $n=3$, it matches the cost of three-party protocols such as~\cite{ara16}. Realizing the dot product operation can also often be performed with the communication cost of a single multiplication, regardless of the size of the input vectors. 

Our computation additionally relies on the following common building blocks:
\begin{itemize}
\item \textbf{Equality.} An equality to zero protocol $[b]_1 \leftarrow {\sf EQZ}([a])$ takes a private integer input $[a]$ and returns a private bit $[b]$, which is set to 1 if $a = 0$ and is 0 otherwise. Equality of private integers $[x]$ and $[y]$ can be computed by calling the protocol on input $[a] = [x] - [y]$. We use a variant of the protocol from~\cite{dam2019} that produces the output bit secret shared over $\mathbb{Z}_2$ (i.e., skips the conversion of the result to the larger ring). 
    
\item \textbf{Comparisons.} $[b] \leftarrow {\sf MSB}([a])$ outputs the most significant bit $[b]$ of its input $[a]$. When working with positive and negative values, $\sf MSB$ computes the sign and is equivalent to the less-than-zero operation. For that reason, the operation can also be used to compare two integers $[x]$ and $[y]$ by supplying their difference as the input into the function. We use the protocol from~\cite{bac2020}.

\item \textbf{Bit decomposition.} $[x_{\ell-1}]_1, \ldots, [x_0]_1 \leftarrow {\sf BitDec}([x], \ell)$ performs bit decomposition an $\ell$-bit input $[x]$ and outputs $\ell$ secret-shared bits. Our implementation uses the protocol from~\cite{dam2019}, with a modification that random bit generation is based on $\sf edaBits$ (see below) and the output bits are secret shared over $\mathbb{Z}_2$ by skipping their conversion to $\mathbb{Z}_{2^k}$. 

\item \textbf{Truncation.} Truncation $[y] \leftarrow {\sf Trunc}([x], \ell, u)$ takes a secret-shared input $[x]$ at most $\ell$ bits long and realizes a right shift by $u$ bits. It outputs $y = \lfloor \frac{x}{2^u} \rfloor$. We invoke this function only on non-negative inputs $x$. Our implementation augments randomized truncation $\sf TruncPr$ from~\cite{bac2020} with $\sf BitLT$ implemented using a generic carry propagation mechanism.



\item \textbf{Prefix AND.} On input $[x_1]_1, \ldots, [x_n]_1$, $\sf PrefixAND$ outputs $[y_1]_1, \ldots, [y_n]_1$, where $y_i = \prod_{j=1}^i x_j$. This is the same as $y_i = \bigwedge_{j=1}^i x_j$ when $x_i$s are binary.
$\sf PrefixAND$ can be realized as described in~\cite{cat2010i} using a generic prefix operation procedure (when operating over a ring). As the inputs are bits, for performance reasons this protocol is carried out in $\mathbb{Z}_2$. 

\item \textbf{Prefix OR.} Protocol $[y_1]_1, \ldots, [y_n]_1 \leftarrow {\sf PrefixOR}([x_1]_1, \ldots$, $[x_n]_1)$ produces $y_i = \bigvee_{j-1}^i x_j$. This operation can also be implemented using a generic prefix operation mechanism and executed over $\mathbb{Z}_2$.

\item \textbf{All OR.} $[y_0]_1, \ldots, [y_{2^n-1}] _1\leftarrow {\sf AllOr}([x_{n-1}]_1, \ldots, [x_0]_1)$ takes $n$ bits and produces $2^n$ bits $y_j$ of the form $\bigvee_{i=0}^{n-1} c_i$, where each $c_i$ is either $x_i$ or its complement $\neg x_i$ and the protocol enumerates all possible combinations. 
The important property is that only one element at position $x = \prod_{i=0}^{n-1} 2^i x_i$ in the output array will be set to 1, while the remaining elements will be 0. The protocol is described in~\cite{bla20}, which we implement over a ring.

\item \textbf{Random bit generation.} Generation of random bits is a lower-level component of many common building blocks including comparisons, bit decomposition, etc. In this work, we use ${\sf edaBit}$ from~\cite{edabits} for this purpose. The protocol $[r], [r_{n-1}]_1, \ldots, [r_0]_1 \leftarrow {\sf edaBit}(n)$ produces random bits $[r_i]$ shared in $\mathbb{Z}_2$ and the integer they represent $r = \prod_{i=0}^{n-1} 2^i r_i$ in $\mathbb{Z}_{2^k}$. 

\item \textbf{Share reconstruction.} Another lower-level protocol on which we rely is $x = {\sf Open}([x], \ell)$ for reconstructing a secret-shared value to the computation participants. To achieve security guarantees, we use a variant that reconstructs $x \in \mathbb{Z}_{2^\ell}$ from $[x]$ where $\ell \le k$. This is achieved by reducing each share modulo $2^\ell$ prior to the reconstruction to guarantee that no information beyond the $\ell$ bits is exchanged during the reconstruction.

\item \textbf{Ring conversion.} $[x]_{k'} \leftarrow {\sf Convert}([x]_k, k, k')$ starts with $x$ secret-shared over $\mathbb{Z}_{2^{k}}$ and produces shares of the same value secret-shared over $\mathbb{Z}_{2^{k'}}$, where $k' > k$, i.e., the target ring is larger. We use the $\sf Convert$ protocol from~\cite{bac2020}.
\end{itemize}
We also develop several other building blocks as described in Section~\ref{sec:prots}. Note that many of these building blocks can be implemented using different variants, where the mechanism for random bit generation plays a particular role. Using the $\sf edaBit$ approach as described above lowers communication cost of protocols compared to generating each random bit separately with shares in $\mathbb{Z}_{2^k}$, but incurs a higher number of communication rounds. We make design choices in favor of lowering communication, but the alternative is attractive when summing a small number of inputs or when the latency between the computational nodes is high.

Notation $\leftarrow$ is used for functionalities that draw randomness (to produce randomized output or to compute a deterministic functionality that internally uses randomization) and notation $=$ is used for deterministic computation.

%% file: sections/large-precision-construction.tex
\section{Secure Large-Precision Construction}
\label{sec:prots}

We are now ready to proceed with our solution for secure and accurate floating-point number summation based on the superaccumulator structure of Section~\ref{sec:superaccu}. 
As before, a floating-point number $x_i$ is represented as a tuple $\langle b_i, v_i, p_i \rangle$. 
Our solution takes a sequence of $n$ secret-shared floating-point inputs $\langle [b_i], [v_i], [p_i] \rangle$ and produces a secret-shared floating-point sum. 
At high level, it proceeds by first converting the inputs into superaccumulators, then computing the sum of the superaccumulators, regularizing the result, and converting the resulting superaccumulator to a floating-point number. The protocol, denoted as $\sf FLSum$, is given in Algorithm~\ref{prot:FLSUMLarge} (superaccumulator summation and regularization are combined into $\sf SASum$). 
\begin{algorithm}[t]
  \caption{$[s] \leftarrow {\sf FLSum}(\langle[b_1], [v_1], [p_1]\rangle, \ldots, \langle[b_{n}], [v_{n}], [p_{n}]\rangle)$}
  \label{prot:FLSUMLarge}
  \begin{algorithmic}[1]
    \STATE let $\alpha = \lceil \frac{2^e+m}{w}\rceil$ and $\beta = \lceil \frac{m+1}{w} \rceil + 1 $;
    \FOR{$i = 1, \ldots, n$ in parallel}
        \STATE $\langle[y_{i, \alpha}], \ldots, [y_{i,1}]\rangle \leftarrow {\sf FL2SA}([b_i], [v_i], [p_i], \alpha, \beta)$;
    \ENDFOR
    \STATE $\langle[y_{\alpha}], \ldots, [y_{1}]\rangle \leftarrow {\sf SASum}(\langle [y_{1, \alpha}], \ldots, [y_{1, 1}] \rangle, \ldots, \langle [y_{n, \alpha}], \ldots$, $[y_{n, 1}] \rangle)$;
    \STATE $\langle[b], [v], [p]\rangle \leftarrow {\sf SA2FL}([y_{\alpha}], \ldots, [y_{1}])$;
    \RETURN $\langle[b], [v], [p]\rangle$; 
  \end{algorithmic}
\end{algorithm}
Data representation parameters $e$, $m$, and $w$ are fixed throughout the computation (as given in Equation~\ref{eq:float}) and are implicit inputs.

When constructing a privacy-preserving solution, the computation that we perform must be data-independent or data-oblivious, as not to disclose any information about the underlying values. In the context of working with the superaccumulator representation, we need to be accessing all superaccumulator slots in the same way regardless of where the relevant data might be located. In particular, when converting a floating-point value to a superaccumulator, at most $\beta$ slots will contain non-zero values, but their location cannot be disclosed. Similarly, when converting a regularized superaccumulator corresponding to the sum to its floating-point representation, only most significant non-zero slots are of relevance, but we need to hide their position within the superaccumulator. 

It is important to note that, unless specified otherwise, the computation is performed over $2w$-bit shares (or ring $\mathbb{Z}_{2^{2w}}$ in our implementation) to facilitate superaccumulator operations. We denote the default element bitlength by $k$. This default bitlength is sufficient to represent all values with a single exception: the bitlength $m$ mantissa $v$ in the floating-point representation can often exceed the value of $2w$. For that reason, we represent mantissa $v$ as as a sequence of $\lceil \frac{m+1}{w} \rceil$, or $\beta-1$, secret-shared blocks storing $w$ bits of $v$ per block. For clarity of exposition, each $v_i$ is written as a single shared value in $\sf FLSum$, while in the more detailed protocols that follow we make this representation explicit.  

For most protocols in this paper, including $\sf FLSum$ in Algorithm~\ref{prot:FLSUMLarge}, security follows as a straightforward composition of the building blocks assuming that the sub-protocols are themselves secure. Then using a standard definition of security that requires a simulator without access to private data to produce corrupt parties' view indistinguishable from the protocol's real execution, we can invoke the simulators corresponding to the sub-protocols and obtain security of the overall construction. Thus, in the remainder of this work we discuss security of a specific protocol only when demonstrating its security involves going beyond a simple composition of its sub-protocols. In addition, for some protocols it is important to ensure that they are data-oblivious (i.e., data-independent) such that the executed instructions and accessed memory locations are independent of private inputs. Data obliviousness is necessary for achieving security because we need the ability to simulate corrupt parties' view without access to private data.

\subsection{Floating-Point to Superaccumulator Conversion}

The first component is to convert floating-point inputs to their superaccumulator representation. Because this operation is rather complex and needs to be performed for each input, it dominates the cost of the overall summation and thus it is important to optimize the corresponding computation. The conversion procedure takes a floating point value $([b], \langle [v_{\beta-1}], \ldots, [v_1] \rangle, [p])$ representing normalized $x = (-1)^b \cdot (1 + 2^{-m}v) \cdot 2^{p-2^{e-1}-1}$ and needs to produce a regularized superaccumulator as a vector of $\alpha$ $2w$-bit integers, where $\alpha = \lceil \frac{2^e+m}{w}\rceil$. 

\subsubsection{The Overall Construction}

To perform the conversion, the computation needs to determine the position within the superaccumulator where the mantissa is to be written based on exponent $[p]$, represent the mantissa as $\beta$ superaccumulator blocks, and write the blocks in the right locations without disclosing what locations within the superaccumulator those are. The protocol details are given as protocol $\sf FL2SA$ (Algorithm~\ref{prot:FL2SA}), which we consequently explain.
\begin{algorithm}[t]
  \caption{$\langle [y_{\alpha}], \ldots, [y_1] \rangle \leftarrow {\sf FL2SA}([b], \langle [v_{\beta-1}], \ldots, [v_1] \rangle$, $[p], \alpha, \beta)$}
  \label{prot:FL2SA}
  \begin{algorithmic}[1]
     \STATE $[p^\mathit{high}] \leftarrow {\sf Trunc}([p], e, \log w)$; 
    \STATE $[p^\mathit{low}] = [p] - [p^\mathit{high}] \cdot w$; 
    \STATE $[z]_1 \leftarrow {\sf EQZ}([p])$;
    \STATE $[v_{\beta-1}] = [v_{\beta-1}] + 2^{m-w(\beta-2)} \cdot {\sf B2A}(1 - [z]_1, 2w)$; 
    \STATE $\langle [v_\beta], \ldots, [v_1] \rangle \leftarrow {\sf Shift}(\langle [v_{\beta-1}], \ldots, [v_1] \rangle,$ $[p^\mathit{low}], w)$; 
     \FOR{$i = 1, \ldots, \beta$ in parallel}
        \STATE $[v_{i}] \leftarrow ([1] - 2 \cdot [b]) \cdot [v_{i}]$; 
    \ENDFOR
   \STATE $\langle [d_{\alpha}], \ldots, [d_1]\rangle \leftarrow {\sf B2U}([p^\mathit{high}]+1, \alpha)$;
    \FOR{$i = 1, \ldots, \alpha$ in parallel} 
        \IF{$ i < \beta$}
             \STATE $[y_i] \leftarrow \sum^{i}_{j=0} [d_{i-j}] \cdot [v_{j}]$;
        \ELSIF{$i \le \alpha - \beta + 1$}
             \STATE $[y_i] \leftarrow \sum^{\beta-1}_{j=0} [d_{i-j}] \cdot [v_j]$;
        \ELSE
             \STATE $[y_i] \leftarrow \sum^{\alpha - 1 - i}_{j=0} [d_{i - \beta + 1 + j}] \cdot [v_{\beta - 1 - j}]$;
        \ENDIF
    \ENDFOR
    \RETURN $\langle[y_{\alpha}], \ldots, [y_1]\rangle$;  
  \end{algorithmic}
\end{algorithm}

Recall that the superaccumulator's step is $2^w$. This means that $e-\log w$ most significant bits of the exponent $[p]$ represent the index of the first non-zero slot in the accumulator. The $\log w$ least significant bits of the exponent are used to shift the mantissa so that it is aligned with the block representation of the superaccumulator. Thus, in the beginning of $\sf FL2SA$ we divide the exponent $[p]$ into two parts: the most significant $e-\log w$ are denoted by $p^\mathit{high}$ and the remaining $\log w$ bits are denoted by $p^\mathit{low}$ (lines 1--2).

The next task is to use the mantissa (represented as $\beta-1$ blocks) and $[p^\mathit{low}]$ to generate $\beta$ superaccumulator blocks. First, recall that normalized floating-point representation assumes that the most significant bit of the mantissa is 1 and is implicit in the floating-point representation. Thus, we need to prepend 1 as the $(m+1)$st bit of $v$. In $\sf FL2SA$ we do this conditionally only when the exponent is non-zero (lines 3--4) because when $p=0$, normalization might not be possible (e.g., if the floating-point value represents a zero). Second, we need to shift the updated mantissa blocks by a private $\log w$-bit value $p^{low}$ to be aligned with the boundaries of superaccumulator blocks and update each value to be $w$ bits by carrying the overflow into the next block. 

\begin{figure*}[t]
\centering
\resizebox{4.8in}{!}{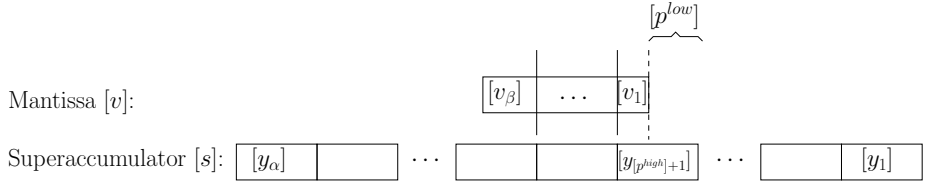} 
\caption{Illustration of floating-point to superaccumulator conversion.} \label{fig:fp2sa}
\end{figure*}

To perform re-partitioning, we considered 
solutions based on bit decomposition and truncation for re-partitioning the blocks, and the second approach was determined to be faster. Our final solution -- a protocol called $\sf Shift$ that takes the original mantissa blocks -- left shifts the values by private $[p^\mathit{low}]$ positions, where $w$ is the upper bound on the amount of shift, and re-aligns the blocks to contain $w$ bits each using truncation. The details of the $\sf Shift$ protocol are deferred to the next sub-section. 
After producing the superaccumulator blocks (line 5), we update the sign of each block using bit $[b]$ (lines 6--8). The desired superaccumulator representation is depicted in Figure~\ref{fig:fp2sa}, where the produced superaccumulator blocks are intended to be written in positions $p^\mathit{high}$ + 1 through $p^\mathit{high}+\beta$.  

The last task is to write the generated $\beta$ superaccumulator blocks $[v_i]$s into the right positions of our $\alpha$-block superaccumulator, as specified by the value of $[p^\mathit{high}]$. Because the computation must be data-oblivious, the location of writing cannot be revealed and the access pattern must be the same for any value of $p^\mathit{high}$. To accomplish the task, we considered two possible solutions: (i) turning the value of $p^{high}$ into a bit array of size $\alpha$ with the $p^\mathit{high}$ value set to 1 and all others set to 0 and using the bit array to create superaccumulator blocks and (ii) creating a bit array with a single 1 in the first location and rotating the bit array by a private amount $p^\mathit{high}$. The first approach was determined to be faster and we describe it next. 

The conversion of $[p^\mathit{high}]+1$, the value of which ranges between 1 and $\alpha$, to a bit array of private bits with the ($p^\mathit{high}$+1)th bit set to 1 can be viewed as binary to unary conversion, denoted by $\sf B2U$. Prior work considered this building block, and specifically in the context of secure floating-point computation~\cite{ali2013}, but prior implementations were over a field. Because computation over a ring of the form $\mathbb{Z}_{2^k}$ can be substantially faster, we design a new protocol suitable over a ring using recent results, as described later in this section. After the binary-to-unary conversion of $p^\mathit{high}+1$ (line 9 of $\sf FL2SA$), each slot of the superaccumulator $[y_i]$ is computed as the dot product of the previously computed data blocks $[v_i]$s and at most $\beta$ bits $[d_j]$s (lines 10--18) because the data blocks need to be written at positions $p^\mathit{high} - \beta + 1$ through $p^\mathit{high}$. 
In particular, for the middle superaccumulator blocks, there are $\beta$ bits and data blocks to consider when creating each superaccumulator block $[y_i]$, while the boundary blocks would iterate over fewer options.
For example, the block $[y_1]$ will be updated to $[v_1]$ only if $[d_{1}] = [1]$, while block $[y_{2}]$ will be updated to $[v_{1}]$ or $[v_{2}]$ in the case of $[d_{1}] = [1]$ or $[d_{2}] = [1]$, respectively. All superaccumulator blocks are updated in parallel with communication cost equivalent to that of $\alpha$ multiplications.

\subsubsection{New Building Blocks.}

What remains is to describe our $\sf Shift$ and $\sf B2U$ protocols.
The $\sf Shift$ protocol takes an integer value (mantissa in the context of this work) stored in $\beta-1$ blocks $[v_{\beta-1}], \ldots, [v_1]$, shifts the value left by a private amount specified by the second argument $[p]$, where the value of $p$ ranges between 0 and $w$ specified by the third argument, and outputs $\beta$ new blocks $[v_\beta], \ldots, [v_1]$. It is implicit in the interface specification that each original block representation has (at least) $w$ unused bits, so that the content of each block can be shifted by up to $w$ positions without losing information. In particular, we assume that each block has $w$ bits occupied, so that after the shift the intermediate result can grow to $2w$ bits before being reorganized to occupy $w$ bits per block.

\begin{algorithm}[t]
  \caption{$\langle [v_{\beta}], \ldots, [v_1] \rangle \leftarrow {\sf Shift}(\langle [v_{\beta-1}], \ldots, [v_1] \rangle, [p], w)$}
  \label{prot:Shift} 
  \begin{algorithmic}[1]
    \STATE let $\gamma = \log w$;
    \STATE $\langle [p_{\gamma}]_1, \ldots, [p_1]_1\rangle \leftarrow {\sf BitDec}([p], \gamma)$;
    \FOR{$j=1, \ldots, \gamma$ in parallel}
    \STATE $[p_j] \leftarrow {\sf B2A}([p_j]_1)$;
    \ENDFOR
    \STATE $[s] \leftarrow \prod^{\gamma}_{j=1} (2^{2^{j-1}} [p_j] + 1 - [p_j])$; 
    \FOR{$i=1, \ldots, \beta-1$ in parallel}
    \STATE $[u_i] \leftarrow [v_i] [s]$; 
    \STATE $[d_i] \leftarrow {\sf Trunc}([u_i], 2w, w)$;
    \ENDFOR
    \FOR{$i=2, \ldots, \beta-1$ in parallel}
        \STATE $[v_i] = [u_i] - 2^w [d_i] + [d_{i-1}]$;
    \ENDFOR
    \STATE $[v_1] = [u_1] - 2^w [d_1]$;
    \STATE $[v_\beta] = [d_{\beta-1}]$;
    \RETURN $\langle [v_\beta], \ldots, [v_1] \rangle$; 
  \end{algorithmic}
\end{algorithm}

The computation, given in Algorithm~\ref{prot:Shift}, starts by bit-decomposing the private amount of shift $[p]$ and converting the resulting bits to ring elements (lines 1--5). The content of each block $[v_i]$ is shifted left (as multiplication by a power of 2) by the appropriate number of positions depending on the value of each bit of the amount of shift: when bit $[p_j]$ is 0, the value is multiplied by 1; otherwise it is multiplied by a power of 2 that depends on the index $j$ (lines 6--8). We then truncate each shifted block (line 9) to split the value into the least significant $w$ bits that the block will retain and the most significant $w$ bits which are the carry for the next block. Each block is consequently updated by taking the carry from the prior block and keeping its $w$ least significant bits (lines 11--15). Because we shift all blocks in the same way, this operation corresponds to a shift with block re-aligning on the boundary of $w$ bits per block. 

Our ring-based solution for binary-to-unary conversion $\sf B2U$ takes a private integer $[a]$ and public range, where $0 < a \le \ell$, and produces a bit array $\langle[b_1], \ldots, [b_{\ell}] \rangle$ with the $a$th bit set to 1 and all other bits set to 0.
Our goal is to have a variant suitable for computation over ring $\mathbb{Z}_{2^k}$ using most efficient currently available tools. Our solution, shown as Algorithm~\ref{prot:b2uRing}, is based on ideas used for retrieving an element of an array at a private index in~\cite{bla20}.
\begin{algorithm}[t]
  \caption{$\langle[b_{1}], \ldots, [b_{\ell}]\rangle \leftarrow {\sf B2U}([a], \ell)$} \label{prot:b2uRing}
  \begin{algorithmic}[1]
    \STATE $q = \lceil \log \ell \rceil$;
    \STATE $[r], [r_{q-1}]_1, \ldots, [r_{0}]_1 \leftarrow {\sf edaBit}(q)$;
    \STATE $\langle [d_{0}]_1, \ldots, [d_{2^q-1}]_1 \rangle \leftarrow {\sf AllOr}([r_{q-1}]_1, \ldots, [r_{0}]_1)$;
    \STATE $c = {\sf Open}([a] - 1 + [r], q)$;
    \FOR{$i = 0, \ldots, \ell-1$ in parallel}
        \STATE $[b_{i+1}] \leftarrow {\sf B2A}(1 - [d_{(c-i)\mod 2^q}]_1)$;
    \ENDFOR
    \RETURN $\langle [b_{1}], \ldots, [b_{\ell}]\rangle$; 
  \end{algorithmic}
\end{algorithm}

The high-level idea consists of generating $\lceil \log \ell \rceil$ random bits $[r_i]$ that collectively represent a random $\lceil \log \ell \rceil$-bit integer $[r]$, generating $\lceil \log \ell \rceil$-ary ORs of $[r]-i$ for all $\log \ell$-bit $i$ and flipping the resulting bits. This creates a bit array with all values set to 0 except the element at private location $[r]$ set to 1. The ORs are computed simultaneously for all values using protocol $\sf AllOr$. Consequently, the algorithm opens the value of $c = r + a$ (modulo $2^{\lceil \log \ell \rceil}$) and uses the disclosed value to position the only 1 bit of the array in location $a$ (i.e., the bit will be set at position $i$ for which $c - i = r + a - i = r$).

Note that the protocol explicitly calls $\sf edaBit$ for random bit generation (and inherits its properties) and there are alternatives. We enhance performance by carrying out the most time-consuming portion of the computation, namely $\sf AllOr$, over a small ring $\mathbb{Z}_2$ because the computation uses Boolean values. 
This means that after producing $2^{\lceil \log \ell \rceil}$ bits through a sequence of calls to $\sf edaBit$, $\sf AllOr$, and $\sf Open$ and array rotation, we need to convert their shares from $\mathbb{Z}_2$ to $\mathbb{Z}_{2^k}$, which we do using binary-to-arithmetic share conversion $\sf B2A$ (line 6). In addition, reconstruction of $c = r + a$ on line 4 needs to be performed using $q$-bit shares to enforce modulo reduction and prevent information leakage, where share truncation prior to the reconstruction is performed by $\sf Open$ itself using the modulus specified as the second argument.

As far as security goes, we note that besides composing sub-protocols the protocol also reconstructs a value which is a function of private input $[a]$ on line 8. Security is still achieved because $[r]$ is a private value uniformly distributed in $\mathbb{Z}_{2^q}$. Thus, the value of $[a]$ is perfectly protected and the opened element of $\mathbb{Z}_{2^q}$ is also uniformly distributed over the entire range. This means that the view is easily simulatable by choosing a random element of $\mathbb{Z}_{2^q}$ as the output of $\sf Open$ and getting the parties to reconstruct that value.

The last component that we would like to discuss is the $\sf B2A$ protocol. Solutions for converting a bit $b$ secret shared over $\mathbb{Z}_2$ to the same value secret shared over larger ring $\mathbb{Z}_{2^k}$, $[b]_k \leftarrow {\sf B2A}([b]_1, k)$, appear in the literature. 
Conventional solutions that use square root computation to generate a random bit (e.g., \cite{dam2019}) temporarily increase the ring to be $\mathbb{Z}_{2^{k+2}}$ for computing intermediate results. In the context of this work, this effectively doubles the size of the ring elements during the computation when we use a ring $\mathbb{Z}_{2^{32}}$ or $\mathbb{Z}_{2^{64}}$. When the number of participants is not large, an alternative is to cast each local share in $\mathbb{Z}_2$ as a share in $\mathbb{Z}_{2^k}$ and have the parties compute XOR of those values over $\mathbb{Z}_{2^k}$. This approach is used in Araki et al.~\cite{ara18} in the three-party setting with honest majority based on replicated secret sharing (RSS) that costs two consecutive multiplications. The approach of Mohassel and Rindal~\cite{moh18} would also require the same communication in two rounds, but the use of the three-party OT procedure in that work reduces the number of rounds to one.

\begin{table}[t] \centering
\begin{tabular}{|c|c|c|} \hline
Protocol & Communication & Rounds \\ \hline
\cite{dam2019} & $6(k+2)$ & 2 \\ \hline
\cite{ara18} & $6k$ & 2 \\ \hline
\cite{moh18} & $6k$ & 1 \\ \hline
Ours & $3k$ & 2 \\ \hline
\end{tabular}
\caption{Comparison of three-party $\sf B2A$ protocols in the honest majority setting with target ring $\mathbb{Z}_{2^k}$. Total protocol communication is reported in bits.}
\label{tab:b2a}
\end{table}
In this work, we design a new solution in the three-party setting using RSS that does not increase the ring size and lowers the cost of prior protocols as illustrated in Table~\ref{tab:b2a}. Unlike many protocols in this work that can be adapted to different types of underlying arithmetic and the number of computational parties, this is the only protocol that specifically uses RSS with $N=3$ and threshold $t=1$.

With RSS when $N=3$, there are three shares representing any secret-shared $[x]$ which we denote as $[x]^{(1)}$, $[x]^{(2)}$, and $[x]^{(3)}$. Each computational party $P_i$ holds two shares with indices different from $i$. For example, $P_2$ holds shares $[x]^{(1)}$ and $[x]^{(3)}$. We use notation $[x]^{(i)}_k$ to denote a share in ring $\mathbb{Z}_{2^k}$. In addition, each participant with access to shares indexed by $i$ holds a (sufficiently long) key ${\sf key}_i$ used as the seed to a pseudorandom generation. For clarity, we refer to a PRG keyed by ${\sf key}_i$ as ${\sf G}_{i}$. A call to ${\sf G}_i.{\sf next}$ produces a pseudorandom ring element.

The input to $\sf B2A$ is a bit secret-shared over $\mathbb{Z}_2$ and we need to convert the bit to the shares over $\mathbb{Z}_{2^k}$ as specified by the second argument. The protocol is given as Algorithm~\ref{prot:B2A}.
\begin{algorithm}[t]
  \caption{$[x]_k \leftarrow {\sf B2A}([x]_1, k)$}
  \label{prot:B2A}
    \textbf{Setup:} Party $P_i$ holds shares and has access to PRGs $\sf G_j$ with indices  $j \neq i$.\ifCONF\\[0.12in]
\begin{enumerate}
\else\\ \begin{compactenum}
\fi
    \item[1:] \vspace{-0.1in} set $[a]_k = \langle[x]^{(1)}_1, 0, 0\rangle$, $[b]_k = \langle 0, [x]^{(2)}_1, 0\rangle$, and $[c]_k = \langle 0, 0, [x]^{(3)}_1\rangle$;
    \item[2:] evaluate $[s]_k = [a]_k \cdot [b]_k$ as follows:
    \ifCONF \begin{enumerate}
    \else \begin{compactenum} 
    \fi
        \item $P_3$ computes $[s]^{(2)}_k = {\sf G}_2.{\sf next}$, $[s]^{(1)}_k = [a]^{(1)}_k \cdot [b]^{(2)}_k - [s]^{(2)}_k$ (in $\mathbb{Z}_{2^k}$), and sends $[s]^{(1)}_k$ to $P_2$;
        \item $P_2$ sets $[s]^{(1)}_k$ to the received value and $[s]^{(3)}_k = 0$;
        \item $P_1$ computes $[s]^{(2)}_k = {\sf G}_{2}.{\sf next}$ and sets $[s]^{(3)}_k = 0$;
    \ifCONF \end{enumerate}
    \else \end{compactenum} \fi
    \item[3:] $[s]_k = [a]_k + [b]_k - 2[s]_k$;
    \item[4:] evaluate $[u]_k = [s]_k \cdot [c]_k$ as follows:
    \ifCONF \begin{enumerate}
    \else \begin{compactenum} \fi
        \item $P_2$ computes $[u]^{(1)}_k = {\sf G}_{1}.{\sf next}$, $u' = [s]^{(1)}_k \cdot [c]^{(3)}_k - [u]^{(1)}_k$ (in $\mathbb{Z}_{2^k}$), sends $u'$ to $P_1$,
        and computes $[u]^{(3)}_k = u' + {\sf G}_{3}.{\sf next}$ (in $\mathbb{Z}_{2^k}$).
        \item $P_1$ receives $u'$, computes $u'' = {\sf G}_{3}.{\sf next}$, $[u]^{(2)}_k = [s]^{(2)}_k \cdot [c]^{(3)}_1 - u''$, and $[u]^{(3)}_k = u' + u''$ (all computation is in $\mathbb{Z}_{2^k}$), and sends $[u]^{(2)}_k$ to $P_3$;
        \item $P_3$ sets $[u]^{(2)}_k$ to the received value and $[u]^{(1)}_k = {\sf G}_{1}.{\sf next}$.
    \ifCONF \end{enumerate}
    \else \end{compactenum} \fi
    \item[5:] $[x]_k = [s]_k + [c]_k - 2[u]_k$;
    \item[6:] \textbf{return} $[x]_k$
\ifCONF \end{enumerate} 
\else \end{compactenum}
\fi
\end{algorithm}
The high-level idea behind the solution is that $x =  [x]^{(1)}_1 \oplus [x]^{(2)}_1 \oplus [x]^{(3)}_1$ and we use the knowledge of the input shares by the parties to evaluate the two XOR operations in the target ring. 
In particular, we can conceptualize the bit shares $[x]^{(i)}_1$ as secret-shared values over $\mathbb{Z}_{2^k}$ represented as  $[a]_k = \langle [x]^{(1)}_1, 0, 0 \rangle$, $[b]_k = \langle 0, [x]^{(2)}_1, 0 \rangle$, and $[c]_k = \langle 0, 0, [x]^{(3)}_1 \rangle$. If we securely evaluate $[a]_k \oplus [b]_k \oplus [c]_k$, we will obtain secret-shared $[x]_k$ in the desired ring, which could be generically accomplished by two sequential multiplications (i.e., $[a] \oplus [b] = [a] + [b] - 2[a]\cdot[b]$). This is also the logic used in \cite{ara18}.

However, given that our shares of $a$, $b$, and $c$ have a special form, the cost of that computation can be reduced. In particular, a typical implementation of the multiplication operation involves multiplying accessible shares locally and re-sharing the products with other parties using fresh randomization to hide patterns. Because in our case some shares are set to 0, their product will be 0 as well, and no re-sharing is needed.
For example, when computing $[a]_k \cdot [b]_k$, the only contributing term to the product is the product of $[a]^{(1)}_k$ and $[b]^{(2)}$, which is computable by $P_3$ in its entirety. As a result of such optimizations, the communication cost of the overall protocol is one ring element per party.

Referring to Algorithm~\ref{prot:B2A}, as mentioned above, the product of $[a]$ and $[b]$ (step 2) can be computed locally by $P_3$, after which the product is re-shared. The re-sharing uses proper $k$-bit elements to hide information about the product and is split by $P_3$ in two shares to which it has access, namely $[s]^{(1)}$ and $s^{(2)}$. This is similar to the re-sharing in regular multiplication (see, e.g., \cite{bac2020}) and involves $P_3$ communicating a single ring element. 

After turning the product into XOR (line 3), the parties need to compute the product of $[s]_k$ and $[c]_k$, where $[s]_k$ has two non-empty shares ($[s]^{(1)}_k$ and $[s]^{(2)}_k$) and $[c]_k$ has one non-empty share ($[c]^{(3)}_k$). This involves $P_2$ computing the product $[s]^{(1)}_k \cdot [c]^{(3)}_k$ and re-sharing by splitting it into two shares and $P_1$ computing the product $[s]^{(2)}_k \cdot [c]^{(3)}_1$ and also re-sharing it. As described in step 4 of Algorithm~\ref{prot:B2A}, $P_2$'s product is split into $[u]^{(1)}_k$ and value $u'$, which becomes a part of $[u]^{(3)}_k$. Similarly, $P_1$'s product is split into $[u]^{(2)}_k$ and value $u''$, which becomes the second component of $[u]^{(3)}_k$. Both $P_1$ and $P_2$ communicate one ring element each to finish re-sharing and let everyone obtains the shares of the product $u$. The party then finish the computation by turning the product into XOR (line 5). The total communication is equivalent to that of a single multiplication. 

We prove the following result:
\begin{claim}
$\sf B2A$ protocol in Algorithm~\ref{prot:B2A} is 1-private in the semi-honest model in the three-party setting in the presence of a single computationally-bounded corrupt party assuming $\sf G$ is a pseudo-random generator.
\end{claim}
\ifCONF \begin{proof}
\else \noindent \textbf{Proof}
\fi
We prove that our $\sf B2A$ protocol in Algorithm~\ref{prot:B2A}
is secure in the presence of a single corrupt party. We
consider corruption of party $P_1$, $P_2$, and $P_3$ in turn and build a corresponding simulator for each case.

\ifCONF \else \medskip \noindent \fi
\textbf{Party $P_1$ is corrupt.} We first assume that party $P_1$ is corrupt, and build the corresponding simulator $S_1$ to simulate its view in the ideal model. The simulator $S_1$ is constructed as follows:
\begin{itemize}
  \item In step 4(a), $S_1$ draws a uniformly random element $u' \leftarrow \mathbb{Z}_{2^k}$ and sends it to party $P_1$ on behalf of party $P_2$.
  \item In step 4(b), $S_1$ receives $[u]^{(2)}_k$ from $P_1$ on behalf of $P_3$.
\end{itemize}
We next compare the view of $P_1$ that the simulator $S_1$ produces with the view of the corrupt party $P_1$ in the real execution. 
In the beginning of the protocol, $P_1$ holds $[b]_k^{(2)} = [x]_1^{(2)}$ and $[c]_k^{(3)} = [x]_1^{(3)}$ and has access to ${\sf G}_2$ and ${\sf G}_3$.
The simulated view consists of $P_1$ receiving a randomly generated $u'$ in step 4(a), while in the real execution it was computed as $u' = [s]_k^{(1)} \cdot [c]_k^{(1)} - {\sf G}_1.{\sf next}$. Now because $P_1$ does not have access to ${\sf G}_1$, the pseudo-random pad ${\sf G}_1.{\sf next}$ information-theoretically protects the value of the product $[s]_k^{(1)} \cdot [c]_k^{(1)}$. Thus, the value of $u'$ in the real execution is pseudo-random. Then because by definition of a pseudo-random generator its output is indistinguishable from a truly random string of the same size to a computationally-bounded adversary, we obtain that the simulated and real views are indistinguishable. 

\ifCONF \else \medskip \noindent \fi
\textbf{Party $P_2$ is corrupt.} Next, consider the case that party $P_2$ is corrupt. We construct simulator $S_2$ as follows:
\begin{itemize}
  \item In step 2(a), $S_2$ draws a uniformly random $[s]^{(1)}_k \leftarrow \mathbb{Z}_{2^k}$ and sends it to $P_2$ on behalf of party $P_3$.
  \item In step 4(b), $S_2$ receives $u'$ from $P_2$ on behalf of $P_1$.
\end{itemize}
At computation initiation time, $P_2$ holds $[a]_k^{(1)} = [x]_1^{(1)}$ and $[c]_k^{(3)} = [x]_1^{(3)}$ and has access to ${\sf G}_1$ and ${\sf G}_3$.
Similar to the case of corrupt $P_1$, $S_2$ only communicates a random value as $[s]^{(1)}_k$ to $P_2$ in step 2(a). In a real execution, $[s]^{(1)}_k$ is computed as $[a]_k^{(1)} \cdot [b]_k^{(2)} - {\sf G}_2.{\sf next}$, where ${\sf G}_2$ is inaccessible to $P_2$ and thus its output information-theoretically protects the product. Because the PRG's output is computationally indistinguishable from a truly random string to a computationally-bounded adversary, $P_2$'s simulated view is computationally indistinguishable from the view in the real execution.

\ifCONF \else \medskip \noindent \fi
\textbf{Party $P_3$ is corrupt.}
Finally, we construct simulator $S_3$ for the case that party $P_3$ is corrupt:
\begin{itemize}
  \item In step 2(a), $S_3$ receives $[s]^{(1)}_k$ from party $P_3$ on behalf of party $P_2$.
  \item In step 4(b), $S_3$ draws a uniformly random value $[u]^{(2)}_k \leftarrow \mathbb{Z}_{2^k}$ and sends it to $P_3$ on behalf of $P_1$.
\end{itemize}
In the beginning of the computation, $P_3$ has access to $[a]_k^{(1)} = [x]_1^{(1)}$, $[b]_k^{(2)} = [x]_1^{(2)}$, ${\sf G}_1$, and ${\sf G}_2$. It then receives a random $[u]^{(2)}_k$ from $S_3$ in the simulated view, while in the real execution the value is computed as  $u' + u''$, where $u'' = {\sf G}_3.{\sf next}$. Due to security of the PRG, its output is pseudo-random and information-theoretically protects $u'$. We obtain that the value $P_3$ is indistinguishable from a truly random string to a computationally-bounded $P_3$. Thus, we obtain that $P_3$'s views in real execution and simulation are computationally indistinguishable.

We conclude that our $\sf B2A$ protocol is secure in the presence of a single semi-honest adversary.
\ifCONF \end{proof}
\else \hfill $\Box$
\fi
$\sf B2A$ is an important building blocks of many other protocols including truncation, ring conversion, bit decomposition, etc. Thus, the above efficient three-party $\sf B2A$ impact performance of the computation. For that reason, we analyze performance of building blocks and our protocols in the three-party setting using RSS as given in Table~\ref{tab:bb}. Note that we separate input-independent computation that can be pre-computed and the remaining (input-dependent) computation.
\begin{table*}[t!]
\setlength{\tabcolsep}{1ex}
\ifCONF \else \centering \rotatebox{90}{ \begin{minipage}{0.95\textheight} \hspace{-0.3in} \fi
\begin{tabular}{|l|c|c|c|c|} \hline
\multirow{2}{*}{Protocol} & \multicolumn{2}{|c|}{Precomputable} & \multicolumn{2}{|c|}{After precomputation} \\ \cline{2-5}
& Communication & Rounds & Communication & Rounds\\ \hline
$\sf Mult$ & 0 & 0 & $3k$ & 1\\ \hline

${\sf Open}(\ell)$, $\ell \le k$ & 0 & 0 & $3\ell$ & 1 \\ \hline

$\sf B2A$ & 0 & 0 & $3k$ & 2 \\ \hline 

$\sf RandBit$ & $3k$ & 2 & 0 & 0 \\ \hline

${\sf edaBit}(k)$ & $3k\log(k) + 7k$ & $\log(k) + 2$ & 0 & 0\\ \hline

${\sf edaBit}(\ell)$, $\ell < k$ & $3 \ell \log(\ell) + 5\ell + 5k$ & $\log(\ell) + 4$ & 0 & 0\\ \hline

${\sf PrefixOr}(n)$ (in $\mathbb{Z}_2$) & 0 & 0 & $1.5n\log(n)$ & $\log(n)$ \\ \hline

${\sf PrefixAnd}(n)$ (in $\mathbb{Z}_2$) & 0 & 0 & $1.5n\log(n)$ & $\log(n)$ \\ \hline

${\sf MSB}(k)$ & $3k\log(k) + 10k$ & $\log(k) + 2$ & $12k-12$ & $\log(k) + 2$ \\ \hline

${\sf EQZ}(k)$ & $3k\log(k) + 7k$ & $\log(k) + 2$ & $6k - 3$ & $\log(k) + 1$ \\ \hline

${\sf Trunc}(\ell, u)$ & $3k\log(k) + 18k$ & $\log(u) + 3$ & $3k + 3\ell + 6u - 6$ & $\log(u) + 3$\\ \hline

${\sf BitDec}(\ell)$, $\ell < k$ & $3\ell\log(\ell) + 5\ell + 5k$ & $\log(\ell) + 4$ & $3 \ell \log(\ell) + 3\ell$ & $\log(\ell) + 1$\\ \hline

${\sf Convert}(k, k')$ & $3k\log(k) + 7k$ & $\log(k) + 2$& $3k'k + 3k\log(k) + 3k$ & $\log(k) + 3$\\ \hline

\multirow{2}{*}{${\sf Shift}(\beta, w)$} & $(\beta-1)(3k \log(k)+18k) +$ & $\max(\log(\gamma)+4$, & $6(\beta - 1)(2k + w - 1) + $ & \multirow{2}{*}{$\gamma + 2\log(\gamma) + 7$} \\ 
& $3\gamma \log(\gamma) + 5\gamma + 5k$ & $\gamma+3)$ & $3\gamma(k + \log(\gamma) + 1) - 3k$ & \\ \hline

${\sf B2U}(\alpha)$ & $[1.2$--$1.5]3\cdot 2^\delta + 3\delta\log(\delta) + 5\delta + 5k$ & $2\log(\delta) + 4$ & $3\alpha k + 3 \delta$ & 3 \\ \hline

\multirow{3}{*}{${\sf Normalize}(\beta, w)$} & $\beta(3k \log(k) + 7k)+$ & $\max(\log(k) + 2, $ & $3k(\beta l + \beta\log(k) + 2l + \beta - 1)+$ & $2\log(l) + \log(k)+$\\ 
& $6l\log(l)+ 17l$ & $\log(l)+2)$ & $1.5(l-m-2)\log(l-m-2)+$ & $\log(l-m-2)+10$\\ 
& & & $3l(\log(l) + 6) - 12$ & \\ \hline
\end{tabular}
\caption{Performance of protocols in the three-party setting based on RSS using ring $\mathbb{Z}_{2^k}$ (bit-level operations are over $\mathbb{Z}_2$). Protocol parameters affecting performance are listed. Total communication across all parties in bits. $\gamma = \lceil \log(w) \rceil$, $\delta = \lceil \log(\alpha) \rceil$, and $l = w \beta$; $\alpha$, $\beta$, $w$, and $m$ are computation parameters.} \label{tab:bb}
\ifCONF \else \end{minipage} } \fi
\end{table*}

Random bit generation $[r] \leftarrow {\sf RandBit}$ (as used, e.g., in $\sf MSB$) is implemented by using local randomness to generate shares of $[r]_1$ over $\mathbb{Z}_2$ and converting them to the larger ring using $\sf B2A$. We favor the use of $\sf edaBit$ in sub-protocols in place of conventional $\sf RandBit$ random bit generation. This lowers the amount of communication, but increases the number of rounds. 

The cost of $\sf AllOr$ as specified in~\cite{bla20} varies based on the size given as an input. For that reason, in Table~\ref{tab:bb} we list a range of constants for values $\alpha$ used with single and double precision in this work (the smallest $\alpha=9$ with single precision and $w=32$ results in constant 1.5 and the largest $\alpha=132$ with double precision and $w=16$ results in constant 1.2).

\subsection{Superaccumulator Summation}

Once we convert the floating-point inputs into superaccumulators, the next step is to do the summation and regularize the result. This corresponds to the protocol $\sf SASum$ given in Algorithm~\ref{prot:supersum}. 
\begin{algorithm}[t]
  \caption{$\langle[y_{\alpha}], \ldots, [y_1]\rangle \leftarrow {\sf SASum}(\langle [y_{1, \alpha}], \ldots, [y_{1, 1}] \rangle, \ldots$, $\langle [y_{n, \alpha}], \ldots, [y_{n, 1}] \rangle)$}
  \label{prot:supersum}  
  \begin{algorithmic}[1]
    \FOR{$i = 1, \ldots, \alpha$ in parallel}
    \STATE $[s_i] = \sum^{n}_{j=1} [y_{j,i}]$;
    \STATE $[b_i] \leftarrow {\sf MSB}([s_i])$;
    \STATE $[y_i] \leftarrow [s_i] \cdot (2[b_i] - 1) $;
    \STATE $[c_{i+1}] \leftarrow {\sf Trunc}( [y_i], 2w, w)$;
    \STATE $[r_i] = [y_i] - [c_{i+1}] \cdot 2^{w}$;
    \STATE $[y_i] \leftarrow [r_i] \cdot [b_i] + [c_i] \cdot [b_{i-1}]$
    \ENDFOR
    \RETURN $\langle[y_{\alpha}], \ldots, [y_{1}]\rangle$;  
  \end{algorithmic}
\end{algorithm}
The summation of superaccumulators is straightforward, where we sum each superaccumulator block as $[s_i] = \sum^{n}_{i=1} [y_{i,j}]$ for $i=1, \ldots, \alpha$ (line 2). The remaining computation regularizes the resulting superaccumulator.
We first compute the absolute value of each block $y_i$ (lines 3--4) and then split the result into $w$ most significant bits (carry for the next block $[c_{i+1}]$) and $w$ least significant bits ($[r_i]$) using truncation (lines 5--6). The final block value is assembled from the carry of the prior block and the remaining portion of the current block using their corresponding signs (line 7). The carry into block 1 is 0. 


Recall that each superaccumulator block is represented as a $2w$-bit integer and we can add at most $n=2^{w-2}$ inputs without an overflow. If one needs to sum more than $2^{w-2}$ inputs, the computation will proceed in layers, where we first sum accumulators in batches of $2^{w-2}$, regularize the result and then do another layer of summation and regularization to arrive at the final regularized superaccumulator.

\subsection{Superaccumulator to Floating-Point Conversion} 

What remains to discuss is the conversion of the regularized superaccumulator representing the summation to the floating-point representation. To maintain security, our protocols needs to obliviously select $\beta$ superaccumulator blocks starting from the first non-zero block without disclosing the location of the selected blocks. In the event that there are fewer than $\beta$ blocks to extract, the solution will still return $\beta$ blocks. 

The superaccumulator to floating-point conversion protocol $\sf SA2FL$ is given as Algorithm~\ref{prot:SA2FL1} and proceeds as follows. 
\ignore{
\begin{algorithm}[t]
  \caption{$\langle [b], [p], [v] \rangle \leftarrow {\sf SA2FL}([y_{\alpha}], \ldots, [y_1])$}
  \label{prot:SA2FL1}
  \vspace{-0.15in}
  \begin{multicols}{2}
  \begin{algorithmic}[1]
    \FOR{$i = \beta, \ldots, \alpha - 1$ in parallel}
        \STATE $[c_i] = 1 - {\sf EQZ}([y_i])$;
    \ENDFOR
    \STATE $[d_{\alpha - 1}] = [c_{\alpha-1}]$;
    \FOR{$i = \alpha - 2, \ldots, \beta$}
        \STATE $[d_i] = [d_{i+1}] + [c_{i}]$;
    \ENDFOR
    
    \FOR{$i = \beta, \ldots, \alpha - 2$ in parallel}
        \STATE $[c'_i] =1 - {\sf EQZ}([d_i])$;
    \ENDFOR
    \STATE $[c'_{\alpha-1}] = [c_{\alpha-1}]$;
    \FOR{$i = \beta, \ldots, \alpha - 2$ in parallel}
        \STATE $[d'_i] = [c'_i] - [c'_{i+1}]$;
    \ENDFOR
    \STATE $[d'_{\beta - 1}] = [1] - [d_{\beta}]$;
    \STATE $[d'_{\alpha - 1}] = [c_{\alpha - 1}]$;
    \FOR{$i=0, \ldots, \beta-1$ in parallel}
        \STATE $[v'_i] \leftarrow \sum^{\alpha - \beta + i}_{j = i} [d'_{j + \beta-1-i}] \cdot [y_j]$;
        \STATE $[v''_i] \leftarrow {\sf Convert}([v'_i], 2w, w\beta)$;
    \ENDFOR
    \STATE $[v'] = \sum^{\beta-1}_{i=0}[v''_i]\cdot 2^{w\times i}$;
    \STATE $\langle [b], [p'], [v] \rangle \leftarrow {\sf Normalize}([v], w\beta, m)$;
    \STATE $[p] \leftarrow [p'] + \sum^{\alpha - \beta}_{i = 0} [d'_{i+\beta-1}] \cdot i \cdot w$;
    \RETURN $\langle [b], [p], [v] \rangle$;  
  \end{algorithmic}
  \end{multicols}
  \vspace{-0.1in}
\end{algorithm}}
\begin{algorithm}[t]
  \caption{$\langle [b], [v_{\beta-1}], \ldots, [v_1], [p] \rangle \leftarrow {\sf SA2FL}([y_{\alpha}], \ldots, [y_1])$}
  \label{prot:SA2FL1}
  \begin{algorithmic}[1]
    \FOR{$i = \beta, \ldots, \alpha$ in parallel}
        \STATE $[c_i]_1 \leftarrow {\sf EQZ}([y_i])$;
    \ENDFOR
    \STATE $\langle [d_\alpha]_1, \ldots, [d_{\beta+1}]_1 \rangle \leftarrow {\sf PrefixAND}([c_\alpha]_1, \ldots$, $[c_{\beta+1}]_1)$;
    \FOR{$i=\beta, \ldots, \alpha$ in parallel}
    \STATE $[d_i] \leftarrow {\sf B2A}([d_i]_1)$;
    \ENDFOR
    \FOR{$i=\beta+1, \ldots, \alpha-1$ in parallel}
    \STATE $[u_i] = [d_{i+1}] - [d_i]$;
    \ENDFOR
    \STATE $[u_\alpha] = 1 - [d_\alpha]$;
    \STATE $[u_\beta] = [d_{\beta+1}]$;
    \FOR{$i=1, \ldots, \beta$ in parallel}
        \STATE $[v_i] \leftarrow \sum^{\alpha - \beta + i}_{j = i} [u_{j + \beta-1-i}] \cdot [y_j]$;
    \ENDFOR
    \STATE $\langle [b], [v_{\beta-1}], \ldots, [v_1], [p] \rangle \leftarrow {\sf Normalize}([v_\beta]$, $\ldots, [v_1])$;
    \STATE $[p] = [p] + \sum^{\alpha - \beta + 1}_{i = 1} [u_{i+\beta-1}] \cdot i \cdot w$;
    \RETURN $\langle [b], [v_{\beta-1}], \ldots, [v_1], [p] \rangle$;  
  \end{algorithmic}
\end{algorithm}
Let $\sf ind$ denote the (private) index of the first non-zero superaccumulator block. We restrict the value we work with to be in the range $\alpha, \ldots, \beta$ to ensure that we can always extract $\beta$ blocks, i.e., ${\sf ind}=\beta$ even if the first non-zero block has the index smaller than $\beta$. Given a regularized superaccumulator $[y_{\alpha}], \ldots, [y_1]$, we first test each block with the index between $\beta$ and $\alpha$ for equality to zero. Once it is determined which blocks are zero, we need to compute the prefix AND of the computed bits (or, equivalently, the prefix OR of their complements) to determine the first non-zero block. Recall that $\sf PrefixAND$, on input $[x_1], \ldots, [x_n]$ outputs $[y_1], \ldots, [y_n]$, where $y_i = \prod_{j=1}^i x_j$. Also, for performance reasons, we do not convert the resulting bits of equality comparisons to full ring element and instead proceed with prefix computation on bits. 

For prefix AND, we start with the highest index and thus the output will be a sequence of 1s followed by 0s starting from the high indices. The first 0 is the value we want to mark differently from others, indicating the first non-zero block. This is accomplished by computing the difference between two adjacent block values (lines 8--12) and we obtain the first non-zero block marked with 1, while all other blocks are as 0. It is important to note that the $\beta$th block will be marked even if all of the blocks $\alpha, \ldots, \beta$ are 0, because in that case we still need to retrieve $\beta$ blocks with the smallest values, i.e., $\sf ind$ is set to $\beta$ and the actual content of the $\beta$th block is irrelevant.

The next step is to extract $\beta$ blocks starting from the marked block, i.e., using the previously introduced notation, we extract the blocks $[y_{\sf ind}], \ldots, [y_{{\sf ind}-\beta + 1}]$ (lines 13--15). We consequently normalize the block using a sub-protocol $\sf Normalize$ that returns a floating-point representation of the blocks, which is consequently updated on line 17 to modify the exponent according to the position of the extracted blocks in the superaccumulator. 

The next protocol, $\sf Normalize$, corresponds to the conversion of $\beta$ extracted superaccumulator blocks to a normalized floating-point value. As before, each block $[v_i]$ is assumed to contain $w$ bits and we normalize the value by finding the first non-zero bit and creating an $m$-bit mantissa with the $(m+1)$st bit set to 1 and the remaining bits partitioned among the output blocks $[v_{\beta-1}], \ldots, [v_1]$. 
\begin{algorithm}[t]
  \caption{$\langle [b], [v_{\beta-1}], \ldots, [v_1], [p]\rangle \leftarrow {\sf Normalize}([v_\beta], \ldots$, $[v_1])$}
  \label{prot:Normalize}
  \begin{algorithmic}[1]
    \STATE let $l = w\cdot \beta$;
    \FOR{$i=1, \ldots, \beta$ in parallel}
        \STATE $[v_i]_l \leftarrow {\sf Convert}([v], k, l)$;
    \ENDFOR
    \STATE $[s]_{l} = \sum_{i=1}^\beta 2^{w(i-1)} [v_i]_l$;
    \STATE $[b]_{l} \leftarrow 1 - 2 \cdot {\sf MSB}([s]_{l})$;  
    \STATE $[v]_{l} \leftarrow [b]_{l} \cdot [s]_{l}$;
    \STATE $\langle [c_{l-1}]_1, \ldots, [c_0]_1 \rangle \leftarrow {\sf BitDec}([v]_{l}, l)$;
    \STATE $\langle [d_{l-2}]_1, \ldots, [d_{m+1}]_1 \rangle \leftarrow {\sf PrefixOR}([c_{l-2}]_1$, $\ldots, [c_{m+1}]_1)$;
    \FOR{$i=0, \ldots, l-1$ in parallel}
    \STATE $[c_i] \leftarrow {\sf B2A}([c_i]_1)$;
    \ENDFOR
    \FOR{$i=m+1, \ldots, l-2$ in parallel}
    \STATE $[d_i] \leftarrow {\sf B2A}([d_i]_1)$;
    \ENDFOR   
    \STATE $[z_{l-1}] = [d_{l-1}]$;
    \FOR{$i = m+1, \ldots, l-2$ in parallel}
        \STATE $[z_i] = [d_i] - [d_{i+1}]$;
    \ENDFOR
    \STATE $[z_{m}] = 1 - [d_{m+1}]$;
    \FOR{$i=0, \ldots, m-1$ in parallel}
        \STATE $[u_i] \leftarrow \sum^{l - 1 - m + i}_{j = i} [z_{j + m - i}] \cdot [c_j]$;
    \ENDFOR
    \FOR{$i=1, \ldots, \beta-2$}
    \STATE $[v_i] = \sum^{w-1}_{j=0}[u_{j+(i-1) w}] \cdot 2^j$;
    \ENDFOR
    \STATE $[v_{\beta-1}] = \sum_{i=w(\beta-2)}^{m-1} [u_i] \cdot 2^{i-w(\beta-2)}$;
    \STATE $[z_{m}] \leftarrow [z_{m}] \cdot [c_m]$;
    \STATE $[p] = \sum^{l-m-1}_{i=0} i \cdot [z_{i+m}]$
    \RETURN $\langle [b], [v_{\beta-1}], \ldots, [v_1], [p]\rangle $;  
  \end{algorithmic}
\end{algorithm}

\ifCONF \perftables \fi

The protocol is given as Algorithm~\ref{prot:Normalize} and proceeds as follows. The first portion of the computation is concerned with assembling the input blocks as a single integer and consequently determining the first non-zero bit. A complicating factor is that different blocks can have different signs, which makes it non-trivial to work at the level of individual blocks. Therefore, the first step of the computation is to convert the shares of the input blocks from the ring with $k=2w$-bit elements to longer $l = w\beta$-bit elements (lines 2--4). The blocks are consequently added together as $[s]_l$ (line 5) and the absolute value of $[s]_l$ is computed as $[v]_l$ (lines 6--7). We next bit-decompose the computed value (line 8) and from this point on the computation can return to shorter $k$-bit shares, but we additionally optimize the computation to skip immediate conversion of bits to $k$-bit shares and run the next step on bit shares as well. 

Given the bits of the value we need to normalize, we determine the first non-zero bit and grab the next $m$ bits (as the $(m+1)$st bit is 1 and is implicit). If there are fewer than $m+1$ non-zero bits, the value must correspond to the lowest blocks of the superaccumulator (as otherwise, the $w\beta$ bits are guaranteed to contain $m+1$ non-zero bits) and cannot be represented in the properly normalized form. In that case we store the $m$ least significant bits in the mantissa and the floating-point value's exponent will be 0. Thus, we first call the prefix OR operation on the most significant $l-m-2$ bits (line 9) and compute the difference between the adjacent bits. As a result, the most significant non-zero bit of $v$ will be set to 1 in $[z_i]$s, with all others set to 0 (lines 16--19). If the first non-zero bit is at position $m$ (when counting from 0) or a lower index, $z_m$ is set to 1 to permit retrieval of $m$ least significant bits (line 20). Then the $m$ bits after the marked bit are extracted (lines 21--23) and are stored in $\beta-1$ blocks (lines 24--27). 

What remains is to form the exponent based on the position of the first non-zero bit. This time we need to distinguish between normalized $(m+1)$-bit mantissas that start from position $m$ and mantissas with fewer than $m+1$ non-zero bits. For that reason, we update the bit $[z_m]$ (line 28) prior to computing the exponent $[p]$ (line 29).

%% file: fig/sc.pdf_t
\begin{picture}(0,0)%
\includegraphics{./fig/sc.pdf}%
\end{picture}%
\setlength{\unitlength}{3947sp}%
\begingroup\makeatletter\ifx\SetFigFont\undefined%
\gdef\SetFigFont#1#2#3#4#5{%
  \reset@font\fontsize{#1}{#2pt}%
  \fontfamily{#3}\fontseries{#4}\fontshape{#5}%
  \selectfont}%
\fi\endgroup%
\begin{picture}(15344,2986)(-396,11017)
\put(9801,12389){\makebox(0,0)[lb]{\smash{{\SetFigFont{23}{26.0}{\rmdefault}{\mddefault}{\updefault}{\color[rgb]{0,0,0}$[v_{1}]$}%
}}}}
\put(-370,11264){\makebox(0,0)[lb]{\smash{{\SetFigFont{25}{30.0}{\rmdefault}{\mddefault}{\updefault}{\color[rgb]{0,0,0}Superaccumulator $[s]$:}%
}}}}
\put(-370,12239){\makebox(0,0)[lb]{\smash{{\SetFigFont{25}{30.0}{\rmdefault}{\mddefault}{\updefault}{\color[rgb]{0,0,0}Mantissa $[v]$:}%
}}}}
\put(6376,11264){\makebox(0,0)[lb]{\smash{{\SetFigFont{25}{30.0}{\rmdefault}{\mddefault}{\updefault}{\color[rgb]{0,0,0}$\cdots$}%
}}}}
\put(11476,11264){\makebox(0,0)[lb]{\smash{{\SetFigFont{25}{30.0}{\rmdefault}{\mddefault}{\updefault}{\color[rgb]{0,0,0}$\cdots$}%
}}}}
\put(3700,11264){\makebox(0,0)[lb]{\smash{{\SetFigFont{25}{30.0}{\rmdefault}{\mddefault}{\updefault}{\color[rgb]{0,0,0}$[y_{\alpha}]$}%
}}}}
\put(13900,11264){\makebox(0,0)[lb]{\smash{{\SetFigFont{25}{30.0}{\rmdefault}{\mddefault}{\updefault}{\color[rgb]{0,0,0}$[y_{1}]$}%
}}}}
\put(10351,13664){\makebox(0,0)[lb]{\smash{{\SetFigFont{25}{30.0}{\rmdefault}{\mddefault}{\updefault}{\color[rgb]{0,0,0}$[p^{low}]$}%
}}}}
\put(9830,11264){\makebox(0,0)[lb]{\smash{{\SetFigFont{20}{22.0}{\rmdefault}{\mddefault}{\updefault}{\color[rgb]{0,0,0}$[y_{[p^{high}]+1}]$}%
}}}}
\put(8851,12314){\makebox(0,0)[lb]{\smash{{\SetFigFont{25}{30.0}{\rmdefault}{\mddefault}{\updefault}{\color[rgb]{0,0,0}$\cdots$}%
}}}}
\put(7651,12389){\makebox(0,0)[lb]{\smash{{\SetFigFont{23}{26.0}{\rmdefault}{\mddefault}{\updefault}{\color[rgb]{0,0,0}$[v_{\beta}]$}%
}}}}
\end{picture}%

%% file: sections/performance.tex
\section{Performance Evaluation}

\ifCONF \else \perftables \fi
In this section, we evaluate performance of our construction and compare it to the state-of-the-art secure floating-point summation protocols.
Our implementation is in C++ using RSS over a ring $\mathbb{Z}_{2^k}$ and is available at \cite{fpSumImpl}. We run all experiments in a three-party setting using machines with a 2.1GHz CPU connected by a 1Gbps link with one-way latency of 0.08ms. All experiments are single threaded and are not optimized for round complexity with respect to pre-processing. Instead, randomness generation is performed inline as specified in the protocols and the actual number of rounds in the implementation is higher than what is possible and what is reported in Table~\ref{tab:bb}. 
Each experiment was executed at least 100 times, and the average runtime is reported.

\begin{figure}[t] \centering
\begin{tikzpicture}
\pgfplotsset{every tick label/.append style={font=\small}}

        \begin{semilogyaxis}[
                height = 2.1in,
        xshift=1.8cm,
xlabel={Input size},
label style={font=\normalsize},
xtick={4,8,12,16,20},
xticklabels={$2^4$,$2^{8}$,$2^{12}$,$2^{16}$,$2^{20}$},
xmajorgrids,
ylabel={Runtime (ms)},
        legend style={font=\footnotesize, at={(0.05,0.75)},anchor=west, legend columns=1},
        ymajorgrids=true,     
        xmajorgrids=true,
        grid style=dashed,                  
    ]

\addlegendentry{${\sf Ours}$}
\addplot[mark size=1pt, color= blue, mark=*] coordinates {(4, 0.190359) (6, 0.259597) (8, 0.257536) (10, 0.415979) (12, 1.500407) (14, 3.551884) (16, 11.442960) (18, 45.667727) (20, 182.591368)};

\addlegendentry{\cite{dam2019}}
\addplot[mark size=1pt, color= red, mark=*] coordinates {(4, 0.210067) (6, 0.255776) (8, 0.379101) (10, 0.900000) (12, 3.162766) (14, 11.569871) (16, 55.238550) (18, 216.187640) (20, 858.949926)};
\end{semilogyaxis}
\end{tikzpicture}  
\caption{Performance comparison of {\sf B2A} protocols.} \label{fig:perf-b2a}
\end{figure}
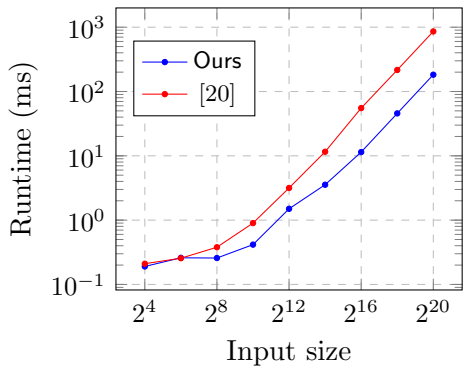

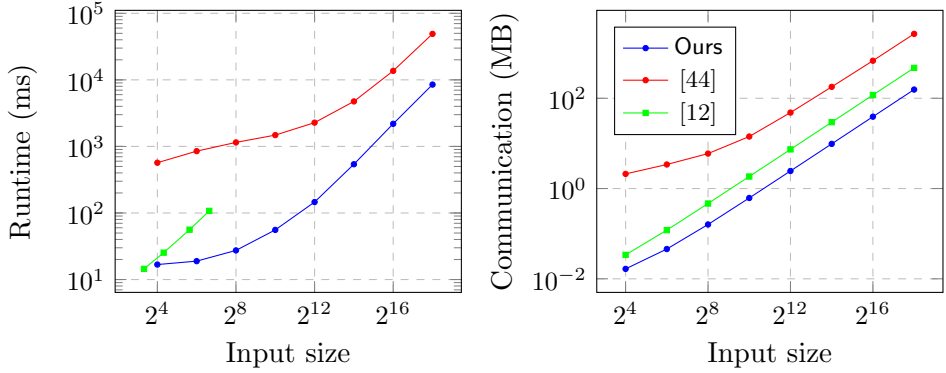
\begin{figure*}[t] \centering
\begin{tikzpicture}
\pgfplotsset{every tick label/.append style={font=\small}}

        \begin{semilogyaxis}[
        height = 2.1in,
        name=ax1,
xlabel={Input size},
label style={font=\normalsize},
xtick={4,8,12,16,20},
xticklabels={$2^4$,$2^{8}$,$2^{12}$,$2^{16}$,$2^{20}$},
xmajorgrids,
ylabel={Runtime (ms)},
        legend style={font=\large, at={(-0.1,0.-0.8)},anchor=north west,legend columns=2},
        ymajorgrids=true,     
        xmajorgrids=true,
        grid style=dashed,                 
    ]      

\addplot[mark size=1pt, color= blue, mark=*] coordinates {(4, 16.8)(6, 18.9)(8, 27.4)(10, 55.7)(12, 146)(14, 540)(16, 2187)(18, 8495)};

\addplot[mark size=1pt, color= red, mark=*] coordinates {(4, 569)(6, 849)(8, 1149)(10, 1479)(12, 2276)(14, 4744)(16, 13648)(18, 49109)};

\addplot[mark size=1pt, color= green, mark=square*] coordinates {(3.321, 14.46)(4.321, 25.34)(5.64, 56.11)(6.64, 107.49)};
\end{semilogyaxis}

        \begin{semilogyaxis}[
                height = 2.1in,
        at={(ax1.south east)},
        xshift=1.8cm,
xlabel={Input size},
label style={font=\normalsize},
xtick={4,8,12,16,20},
xticklabels={$2^4$,$2^{8}$,$2^{12}$,$2^{16}$,$2^{20}$},
xmajorgrids,
ylabel={Communication (MB)},
        legend style={font=\footnotesize, at={(0.05,0.75)},anchor=west, legend columns=1},
        ymajorgrids=true,     
        xmajorgrids=true,
        grid style=dashed,                  
    ]

\addlegendentry{${\sf Ours}$}
\addplot[mark size=1pt, color= blue, mark=*] coordinates {(4, 0.0166)(6, 0.0457)(8, 0.16)(10, 0.618)(12, 2.45)(14, 9.77)(16, 39.07)(18, 156)};

\addlegendentry{\cite{SecFloat}}
\addplot[mark size=1pt, color= red, mark=*] coordinates {(4, 2.11)(6, 3.41)(8, 5.95)(10, 14.2)(12, 48.2)(14, 179)(16, 679)(18, 2673)};

\addlegendentry{\cite{cat2020p}}
\addplot[mark size=1pt, color= green, mark=square*] coordinates {(4, 0.0339)(6, 0.12)(8, 0.466)(10, 1.847)(12, 7.37)(14, 29.5)(16, 118)(18, 471)};

\end{semilogyaxis}
\end{tikzpicture}  
\caption{Performance comparison with related work for single precision. [9]'s runtime uses different hardware.} \label{fig:perf-comp}
\end{figure*}

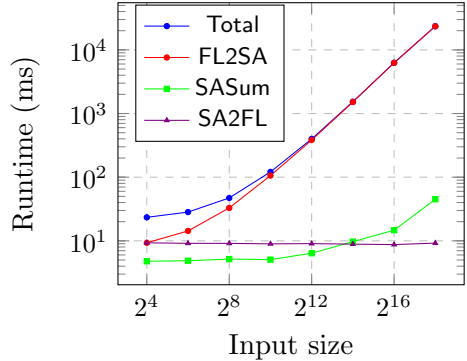
\begin{figure}[t] \centering
\begin{tikzpicture}
\pgfplotsset{every tick label/.append style={font=\small}}

        \begin{semilogyaxis}[
                height = 2.1in,
        xshift=0.1cm,
xlabel={Input size},
label style={font=\normalsize},
xtick={4,8,12,16,20},
xticklabels={$2^4$,$2^{8}$,$2^{12}$,$2^{16}$,$2^{20}$},
xmajorgrids,
ylabel={Runtime (ms)},
        legend style={font=\footnotesize, at={(0.05,0.75)},anchor=west, legend columns=1},
        ymajorgrids=true,     
        xmajorgrids=true,
        grid style=dashed,                  
    ]

\addlegendentry{${\sf Total}$}
\addplot[mark size=1pt, color= blue, mark=*] coordinates {(4, 23.4)(6, 28.3)(8, 47.2)(10, 121)(12, 399)(14, 1536)(16, 6271)(18, 23540)};

\addlegendentry{${\sf FL2SA }$}
\addplot[mark size=1pt, color= red, mark=*] coordinates {(4, 9.32)(6, 14.3)(8, 32.9)(10, 106)(12, 383)(14, 1517)(16, 6247)(18, 23485)};

\addlegendentry{${\sf SASum}$}
\addplot[mark size=1pt, color= green, mark=square*] coordinates {(4, 4.78)(6, 4.87)(8, 5.17)(10, 5.07)(12, 6.39)(14, 9.71)(16, 14.7)(18, 45)};

\addlegendentry{${\sf SA2FL}$}
\addplot[mark size=1pt, color= violet, mark=triangle*] coordinates {(4, 9.31)(6, 9.14)(8, 9.12)(10, 8.97)(12, 9.04)(14, 8.87)(16, 8.73)(18, 9.21)};

\end{semilogyaxis}
\end{tikzpicture}  
\caption{Performance of double precision protocol with $w = 32$.} \label{fig:perf-double}
\end{figure}

To evaluate the impact of our new three-party $\sf B2A$ protocol, in Figure~\ref{fig:perf-b2a} we provide performance comparison of a common  square root based solution from \cite{dam2019} and our solution described in Algorithm~\ref{prot:B2A}. Because the former requires a slightly larger ring size $\mathbb{Z}_{2^{k+2}}$, we set the computation over $\mathbb{Z}_{2^k} = \mathbb{Z}_{2^{60}}$ and thus portion of the computation for the protocol from~\cite{dam2019} are over $\mathbb{Z}_{2^{62}}$. The implications are that both protocols can internally use 64-bit arithmetic and the increase in the ring size does not impact communication in bytes. Therefore, communication and the number of rounds of the protocol from~\cite{dam2019} are also the same as those numbers for the protocol from~\cite{ara18}. Had we chosen $k=32$ or $k=64$, the gap in performance between our protocol and that from~\cite{dam2019} would increase due to the need of the later to increase the communication size and use a longer data type for the computation.

As we see from Figure~\ref{fig:perf-b2a}, for smaller input sizes, both solutions exhibit similar performance due to their equivalent round complexity. However, as input size increases beyond $2^6$ and communication and computation become dominant factors in overall performance, our solution outperforms \cite{dam2019} by a significant margin. For instance, the performance gap between the two approaches is as large as a factor of four for input size $2^{20}$, demonstrating the advantage of our $\sf B2A$ protocol even beyond savings in communication.

Performance of our superaccumulator-based floating-point summation for single and double floating-point precision is provided in Table~\ref{tab:perf}.
The performance is additionally visualized in Figures \ref{fig:perf-comp} and ~\ref{fig:perf-double}.
We see that the bottleneck of the summation for both single and double precision is the conversion ${\sf FL2SA}$, particularly when the input size $n$ is large. This is expected because we need to convert all $n$ inputs into the superaccumulator representation. In contrast, superaccumulator to floating-point conversion ${\sf SA2FL}$ has a constant runtime for all input sizes because we only need to convert a single result and the workload does not change. Although summation ${\sf SASum}$ has communication complexity independent of $n$, its local computation linearly depends on the input size, which makes its runtime increase with $n$.

If we compare the runtimes for different values of $w$, using $w = 16$ results in lower overall runtime with single precision, while $w=32$ is superior for double precision. The difference in performance mainly stems from the impact of the choice of $w$ on the performance of ${\sf FL2SA}$ and its dependence on parameters $\alpha$ and $\beta$ (which $w$ directly influences). 

We also compare performance of our superaccumulator-based solution with floating-point summations from \cite{cat2020, cat2020p, SecFloat}. We execute SecFloat's \cite{SecFloat} pairwise addition in a tree-like manner to realize floating-point summation and measure the performance on our setup. Note that SecFloat is for the two-party setting (dishonest majority) and was implemented only for single precision. 
We also include published runtimes of the best performing solution, SumFL2, from~\cite{cat2020p} as the implementation has not been released.
The experiments in~\cite{cat2020p} were run using three 3.6GHz machines connected via a 1Gbps LAN, where the round-trip time (RTT) measured via ping was reported to be 0.35ms (our RTT measured via ping averaged at 0.25ms). We also calculate the communication cost of SumFL2 using the specified formula.\footnote{In~\cite{cat2020p}, communication measured from the implementation differed from communication derived from the analysis and the implementation's communication is 9.3\% lower of the analytical cost. Because the measurement included only one data point with 10 operands, we report results computed according to the formula.} The results are given in Figure~\ref{fig:perf-comp}, where our single-precision solution uses $w=16$.

As shown in the figure, our protocol has better runtime and communication costs than the other two solutions. Although~\cite{SecFloat} states that their implementation is not optimized for batch sizes smaller than $2^{10}$,
our protocol is still 5 times faster and uses 17 times less communication than \cite{SecFloat} with $2^{18}$ inputs. For input sizes larger than $2^{14}$, both solutions demonstrate the same trend. We expect our advantages would be larger in the WAN setting where bandwidth is limited and communication is the bottleneck. 

\begin{table}[t]
  \begin{center}
  \begin{tabular}{|c|c|c|c|c||c|c|c|c|} 
  \hline
  \multirow{3}{*}{Prot.}& \multicolumn{8}{c|}{Input size} \\ \cline{2-9}
  & \multicolumn{4}{c||}{Single} & \multicolumn{4}{c|}{Double} \\ \cline{2-9}
   & 10& 20& 50& 100 & 10& 20& 50& 100\\ \hline 
   Ours &16.5& 17.4& 18.6& 21.2 & 22.3& 24.1& 27.3& 30.5\\ \hline 
   \cite{cat2020p} & 14.5 & 25.3 & 56.1 & 107.5 & 26.5& 43.8 & 95.4 & 158\\ \hline 
  \end{tabular}
  \end{center}
  \caption{Runtime comparison with SumFL2 from~\cite{cat2020p} in ms.}\label{tab:fpcomp}
\end{table}
Compared to \cite{cat2020p}, our best performing configuration has a better runtime despite running on slower machines, as additionally shown in Table~\ref{tab:fpcomp}. In~\cite{cat2020p}, performance is reported with at most 100 inputs.
When $n=100$, our solution demonstrates the largest improvement, being 5 times faster than SumFL2 from~\cite{cat2020p} for both single and double precisions. We expect the improvement to be even larger as the number of inputs increases. 
Furthermore, we note that our solution enjoys higher precision, as the goal of this work was to provide better precision than what is achievable using conventional floating-point addition. Lastly, while~\cite{cat2021} discussed additional optimizations to floating-point polynomial evaluation, it is difficult to extract times that would correspond to the summation.